\documentclass[%
preprint,
 amsmath,amssymb,
 aps,
]{revtex4-2}

\usepackage{float}
\usepackage{graphicx}
\usepackage{dcolumn}
\usepackage{bm}
\graphicspath{{figs/}}

\begin{document}
\newcommand{\scs}{\scriptscriptstyle}
\newcommand{\smallm}{\scs (\!-\!)}
\newcommand{\smallp}{\scs (\!+\!)}
\newcommand{\smallmp}{\scs (\!\mp\!)}
\newcommand{\smallpm}{\scs (\!\pm\!)}

\title{Fano resonances induced by symmetry protected bound states in the continuum in dielectric metasurfaces: line-shape prediction by machine learning method}
\author{V.~S.~Gerasimov$^{1,2}$}
\author{A.~S.~Kostyukov$^{1}$}
\author{A.~E.~Ershov$^{1,2}$}
\author{D.~N.~Maksimov$^{1,3}$}
\author{V.~Kimberg$^{4}$}
\email{victor.kimberg@gmail.com}
\author{M.~S.~Molokeev$^{3,5}$}
\author{S.~P.~Polyutov$^{1}$}

\affiliation{$^1$ International Research Center of Spectroscopy and Quantum Chemistry Siberian Federal University, 660041, Krasnoyarsk, Russia}
\affiliation{$^2$ Institute of Computational Modelling SB RAS,
660036, Krasnoyarsk, Russia}
\affiliation{$^3$ Kirensky Institute of Physics, Federal Research
Center KSC SB RAS, 660036, Krasnoyarsk, Russia}
\affiliation{$^4$ Theoretical Chemistry and Biology, KTH Royal Institute of Technology, Stockholm 106 91, Sweden\\}
\affiliation{$^5$ Laboratory of Theory and Optimization of Chemical and Technological Processes, University of Tyumen, Tyumen 625003, Russia}

\date{\today}

\begin{abstract}

We consider resonances induced by symmetry protected bound states in the continuum in dielectric gratings with in-plane mirror symmetry. It is shown that the shape of the resonance in transmittance is controlled by two parameters in a generic formula which can be derived in the framework of the coupled mode theory. It is numerically demonstrated that the formula encompasses various line-shapes including asymmetric Fano, Lorentzian, and anti-Lorentzian resonances. It is confirmed that the transmittance zeros are always present even in the absence up-down symmetry. At the same time reflectance zeros are not generally present in the single mode approximation. It is found that the line-shapes of Fano resonances can be predicted to a good accuracy by the random forest machine learning method which outperforms the standard least square methods approximation in error by an order of magnitude in error with the training dataset size $N\approx 10^4$.

\end{abstract}

\maketitle
\section{Introduction}
\label{intro}

Optical bound states in the continuum (BICs) are source-free localized solutions of Maxwell's equations which are spectrally embedded into the continuum of scattering states \cite{Hsu16, koshelev2019meta, Koshelev19, joseph2021, Kang2023}. The optical BICs in dielectric metasurfaces have recently become an important
instrument for resonant enhancement of light-matter interaction  to be employed for resonant light absorption~\cite{zhang2015ultrasensitive,wang2020controlling,sang2021highly,xiao2021engineering, cai2022enhancing}, sensing \cite{Liu17,Romano18b}, harmonic  generation~\cite{ndangali2013resonant,  wang2018large, carletti2018giant, koshelev2020subwavelength}, and lasing~\cite{Kodigala17,hwang2021ultralow, Yu2021ultra, yang2021low}. Although BICs are not coupled to the incident light, breaking the system's symmetry under variation of some control parameter ~\cite{Koshelev18, Maksimov20} results  in the so-called qusi-BICs, i.e. long-lived resonant modes with the quality factor diverging to infinity on approach to the BIC in parametric space. This divergence is visible in the transmittance spectrum as a collapsing Fano resonance~\cite{Shipman,SBR,Blanchard16, bogdanov2019bound, pankin2020fano, Bulgakov18b} and simultaneously leads to electromagnetic field enhancement in the host metasurface \cite{Yoon15, Mocella15a}. This picture is generic in nanophotonics since high-quality resonant modes of any kind reveal themselves as sharp Fano resonances in the transmittance spectrum \cite{Campione2016, Zhou2014, Limonov2017, Krasnok19}. As shown in \cite{Fan03} in the single resonant coupled mode approximation the Fano resonances can be described as a product of interference between two optical pathways, namely the resonant pathway due to the excitation of the resonant mode, and the direct or non-resonant pathway due to frequency independent background. 

In this work we investigate the line shape-shapes of Fano resonances induced by symmetry protected optical BICs in dielectric gratings. The symmetry breaking leading to transformation of BICs to quasi-BICs is controlled by small deviation of the angle of incidence from the normal. Our goal is to analyse possible resonant line-shapes and find out weather the line-shapes can be predicted by machine learning methods using the geometric and optical properties as the input parameters. Our tool for describing the Fano resonance line-shape is the themporal coupled mode theory (TCMT) \cite{Fan03}. Nowadays, the TCMT is recognized as an efficient tool for describing the spectra of various phonic devices 
\cite{ Alpeggiani2017, Ming2017, Zhou2016, maksimov2020optical, Bikbaev21, Zhang2023, Wu2022, Huang2024} due to both universality and the clear physical picture it provides. It is worth mentioning that we are going to consider optical systems without up-down symmetry which can affect the line-shape of Fano resonances \cite{Popov1986, Shipman2012, Wang2013, Bykov15, Zhou2016, Yuan2022}. At the same time, machine learning techniques have already been applied to various problems of nanophotonics \cite{ma2021deep, jiang2021deep, so2020deep, pilozzi2018machine, kudyshev2020machine, Zhao2023, Deng2025} including problems related to optical BICs \cite{lin2021engineering, ma2022strategical,wang2023automatic,Wang2023}. Recently, the TCMT approach has been hybridized with neural networks  \cite{Zhang2023, Zhang2024, Su2024} for resonant response synthesis in photonic devices. Here we follow our previous work
 \cite{Molokeev2023} where we showed that the random forest machine learning method is capable of predicting the frequency of optical BICs in symmetric dielectric metasurfaces. In what follows we revisit the TCMT in application to Fano resonances induced by symmetry protected BICs and apply the random forest method in combination with the TCMT to adress the line-shape prediction problem.


\begin{figure}[t]
    \centering
    \includegraphics[width=\textwidth]{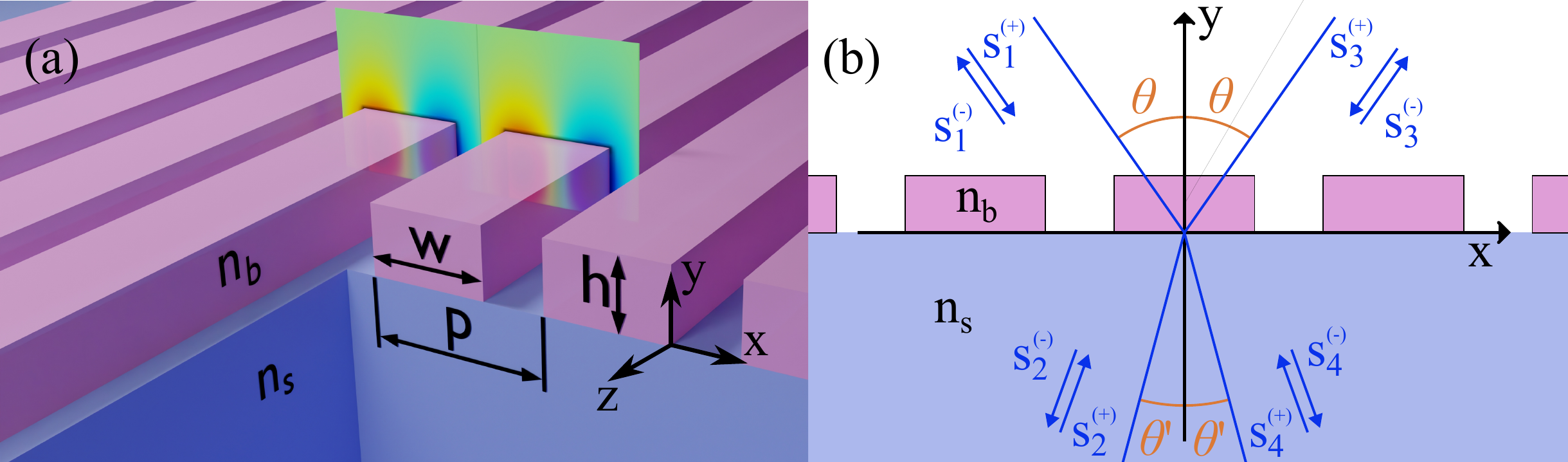}
    \caption{Metasurface in the form of a dielectric grating on a dielectric substrate.  (a) Schematic of the metasurface with the electric field of the antisymmetric BIC mode. (b) Definition of the scattering channels.}
    \label{Fig1}
\end{figure}\section{TCMT equations}

As one can see in Fig.~\ref{Fig1}~(a) the system under scrutiny is a ruled grating made of dielectric bars with refractive index $n_b$. The grating is placed on top of a dielectric substrate with refractive index $n_s$. The supestrate of the system is air with $n_0=1$. All geometric parameters, including the period $p$, the width $w$, and the height $h$ are specified in Fig.~\ref{Fig1}~(a). In what follows we take $p=0.697~\mu{\rm m}$. I this
work we only consider TM-waves which propagate along the $x$-axis but not along the bars, so the scattering problem can be solved in the framework of 2D electrodynamics. The mode profile of a symmetry protected BIC is shown in the Fig.~\ref{Fig1}~(a) in the form of the $z$-component of the electric field.

One can see in Fig.~\ref{Fig1}~(b) that the scattering problem is described by a $4\times 4$ scattering matrix $\widehat{S}_{\scs{4}}$, which links the vectors of incident and outgoing amplitudes as follows
\begin{equation}\label{Smatrix}
{\bf s}_{\smallm}
=\widehat{S}_{\scs{4}}
{\bf s}_{\smallp},
\end{equation}
where the outgoing ${\bf s}^{\smallm}$ and incident ${\bf s}^{\smallp}$ vectors are given by
\begin{equation}
\bf{s}_{\scs{(\pm)}}=
\left(
\begin{array}{c}
s_1^{\scs{(\pm)}} \\
s_2^{\scs{(\pm)}} \\
s_3^{\scs{(\pm)}} \\
s_4^{\scs{(\pm)}}
\end{array}
\right).
\end{equation}
We assume that all dielectric are lossless so the $S$-matrix is unitary $\widehat{S}_{\scs{4}}^{\dagger}\widehat{S}_{\scs{4}}=\widehat{\mathbb I}$. The system also possesses time-reversal symmetry therefore the $S$-matrix is symmetric $\widehat{S}_{\scs{4}}^{\intercal}=\widehat{S}_{\scs{4}}$. Importantly the matrix $\widehat{S}_{\scs{(4)}}$ must be of the block form 
\begin{equation}\label{S4}
    \widehat{S}_{\scs{4}}=
\left(
\begin{array}{cc}
0 & \widehat{S}_{\scs{2}} \\
\widehat{S}_{\scs{2}}^{\intercal} & 0
\end{array}
\right)
\end{equation}
to ensure the momentum conservation in scattering from the metasurface \cite{Zhou2016}, where $\widehat{S}_{\scs{2}}$ is a $2\times 2$ unitary matrix. Since the system has the $\sigma_v$ mirror symmetry the problem can be reduced to finding $\widehat{S}_2$ that is a symmetric matrix as shown in Appendix~\ref{App1}.
Without a loss of generality we can focus on the left-going waves, so that
\begin{equation}\label{S_2}
\left(
\begin{array}{c}
s_{1}^{\smallm} \\
s_{2}^{\smallm} \\
\end{array}
\right)
=\widehat{S}_2
\left(
\begin{array}{c}
s_{3}^{\smallp} \\
s_{4}^{\smallp} \\
\end{array}
\right).
\end{equation}

According to \cite{Fan03} the TCMT equations describing single-mode scattering take the following form
\begin{align}\label{TCMT}
& \frac{d a(t)}{d t}=-(i\omega_0+\gamma)a(t)+{\bf \kappa}^{\intercal}{\bf s^{\smallp}}(t), \nonumber \\
&{\bf s}^{\smallm}(t)=\widehat{C}{\bf s}^{\smallp}(t)+
a(t){\bf d}, 
\end{align}
where $\widehat{C}$ is the matrix of direct (non-resonant) process, $\omega_0$ is the resonance center frequency,
$\gamma$ is the radiation decay rate, $a(t)$ -- the amplitude of the resonant eigenmode, ${\bf \kappa}$ is the coupling vector  and
${\bf d}$ is the decoupling vector. In what follows we assume that the system is illuminated by a monochromatic wave of frequency $\omega$, so all time-dependent quantities in Eq.~\eqref{TCMT} oscillate in time with the harmonic factor $e^{-i\omega t}$. Importantly, the parameters in the TCMT equations are not independent, but linked to each other due to constraints imposed by energy conservation, Lorentz reciprocity and time reversal symmetry~\cite{Fan03, Zhao19}. As we have already mentioned, the energy conservation manifests itself in the unitarity of the S-matrix whereas the time-reversal symmetry forces the S-matrix to be symmetric. 
In this situation the parameters of the TCMT equations are known to satisfy the following three equations  \cite{Fan03}
\begin{align}\label{TCMTcond}
   & 2\gamma={\bf d}^{\dagger}{\bf d}, \nonumber \\
   &    {\bf \kappa}={\bf d}, \nonumber \\
   &    \widehat{C}{\bf d}^{*}+{\bf d}=0.
\end{align}
In our case both energy conservation and time-reversal are present, however, care is needed in application of the time reversal operation since it maps the left-going waves onto the right-going ones. In Appendix~\ref{App2} we show that the $2\times2$ unitary and symmetry of $\widehat{S}_2$ lead to the same constraints for the coupling parameters as
in Eq.~\eqref{TCMTcond}.

Now we have to solve Eq.~\eqref{TCMTcond} for the decoupling vector. We set out from the most generic form of $\widehat{C}$, which is unitary and symmetric,
\begin{equation}\label{Cmatrix}
\widehat{C}
=e^{i\phi}\left(
    \begin{array}{cc}
        \rho e^{-i\eta} & i\tau \\
        i\tau  & \rho e^{i\eta} 
    \end{array}
    \right), \ \rho=\sqrt{1-\tau^2} \\
\end{equation}
with $\tau \in [-1,~1]$, where all parameters are real valued. The number of parameters in Eq.~\eqref{Cmatrix} can be reduced by redefining the incident channels via a unitary transformation
\begin{equation}\label{uni1}
\left(
\begin{array}{c}
     {s}_1^{\smallp} \\
      {s}_2^{\smallp}
\end{array}
\right)=
\left(
\begin{array}{cc}
     e^{i\phi_1} & 0 \\
      0 & e^{i\phi_2}
\end{array}
\right)
\left(
\begin{array}{c}
     \tilde{s}_1^{\smallp} \\
      \tilde{s}_2^{\smallp}
\end{array}
\right).
\end{equation}
To be consistent with the time-reversal symmetry the outgoing channels have to be transformed as follows
\begin{equation}\label{uni2}
\left(
\begin{array}{c}
     {s}_1^{\smallm} \\
      {s}_2^{\smallm}
\end{array}
\right)=
\left(
\begin{array}{cc}
     e^{-i\phi_1} & 0 \\
      0 & e^{-i\phi_2}
\end{array}
\right)
\left(
\begin{array}{c}
     \tilde{s}_1^{\smallm} \\
      \tilde{s}_2^{\smallm}
\end{array}
\right).
\end{equation}
Then, by using Eq.~\eqref{uni1} and Eq.~\eqref{uni2} together with Eq.~\eqref{Cmatrix} one arrives at a single parameter family of unitary symmetric matrices
\begin{equation}\label{Cmatrix_final}
\widehat{C}
=\left(
    \begin{array}{cc}
        \rho  & i\tau \\
        i\tau  & \rho
    \end{array}
    \right), \ \ \rho=\sqrt{1-\tau^2},  \ \ \tau\in[-1, 1],
\end{equation}
if one chooses
\begin{align}
\phi_1=\frac{\eta-\phi}{2}, \
\phi_2=-\frac{\eta+\phi}{2} 
\end{align}
in Eq.~\eqref{uni1}. It can be easely checked that the above unitary transformation complies with Eq.~\eqref{TCMTcond}. The unitary transformation in Eq.~\eqref{uni1} can be thought of as a shifting the reference plane between the scattering domain and the outer space along the $y$-axis. This is always possible in the far-field where the solution is exhaustively described by the scattering channels. By using Eq.~\eqref{Cmatrix_final} in Eq.~\eqref{TCMTcond} one finds a single-parametric family of solutions for $\bf{d}$ as follows
\begin{equation}\label{decoup}
\begin{split}
    {\bf d}=&\sqrt{\frac{\gamma}{(1+\rho)}}
    \left(
    \begin{array}{c}
    \tau \cos\alpha-i(1+\rho) \sin\alpha   \\
        \tau \sin\alpha-i(1+\rho) \cos\alpha 
    \end{array}
    \right), \\ \alpha\in&[-{\pi}/{2}, {\pi}/{2}],
    \end{split}
\end{equation}
The derivation details are presented in Appendix~\ref{App3}.

After using ${\bf \kappa}={\bf d}$ in Eq.~\eqref{0A6} one finds the final solution for the $S$-matrix
\begin{equation}\label{S}
\widehat{S}=\widehat{C}+\frac{{\bf d} {\bf d}^{\intercal}}{i(\omega_0-\omega)+\gamma}.
\end{equation}
The transmission coefficient independent of the direction of incidence is written as
\begin{equation}
\label{TT}
T=\frac{[\tau(\omega_0-\omega)+ \rho\gamma\sin(2\alpha)]^2}
{(\omega_0-\omega)^2+\gamma^2}. 
\end{equation}
If $\alpha=\pm \pi/4$ Eq.~\eqref{TT} limits to the well-known solution presented in \cite{Fan03} for system with up-down symmetry. The transmettance spectrum Eq.~\eqref{TT} complies with the earlier result from \cite{Wang2013} on the fundamental bounds on decay rates in asymmetric single-mode optical resonators, where it was shown that the transmittance is only bound to peak to unity in symmetric resonators.

The system supports a symmetry protected BIC in the $\Gamma$-point. With variation of the angle of incidence
 $\theta$ in the vicinity of the $\Gamma$-point the
BIC is transformed to a high-$Q$ resonant mode with the resonant frequency $\omega_0$ and the decay rate $\gamma$
given by the following Taylor expansion
\begin{align}\label{Taylor1}
& \omega_0=\omega_{\scs \rm {BIC}}+\kappa_{\omega}\theta^2+\mathcal{O}(\theta^4), \nonumber \\
& \gamma=\kappa_{\gamma}\theta^2+\mathcal{O}(\theta^4).
\end{align}
Upon using the Taylor expansion Eq.~\eqref{Taylor1} in Eq.~\eqref{TT} one arrives at
\begin{equation}\label{TTtheta}
T=\frac{[\tau(\omega_{{\scs \mathrm{BIC}}}+\omega_0^{\scs{(2)}}\theta^2-\omega)+ \rho\gamma^{\scs{(2)}}\sin(2\alpha)\theta^2]^2}
{(\omega_{{\scs \mathrm{BIC}}}+\omega_0^{\scs{(2)}}\theta^2-\omega)^2+(\gamma^{\scs{(2)}}\theta^2)^2}+{\cal O}(\theta^6),
\end{equation}
which gives the line-shape of the Fano resonance induced by a symmetry protected BIC.

\section{Dataset acquisition
} 
\label{sec:2}
Our goal is to predict the shapes of the BIC-induced Fano resonance in the system shown in Fig.~\ref{Fig1}. According to Eq.~\eqref{TT}, besides the resonance center-frequency $\omega_0$ and the
radiation decay rate $\gamma$, which are specified by dispersion of leaky band hosting the BIC, there are
only two parameters characterizing the shape of the Fano resonance, namely $\alpha$ and $\tau$.
Both parameters can be found by fitting the numerically computed transmittance
spectrum at the incidence angle slightly different from normal. In this work we take $\theta=2 \ \mathrm{deg}$. 
The radiation decay rate and the center-frequency dictate the position and the width of the Fano resonance, correspondingly. Therefore, for predicting the shape of the resonance we have to analyse how the four parameters $n_1,~n_2,~h$ and $w$ affect the quantities of $\alpha$ and $\tau$. Here we address this problem by applying machine learning algorithm to the data set obtained by numerically solving Maxwell's equation under variation of all four control parameters. The ranges of parameters are specified below
\begin{align}\label{range}
n_1\in[2, 5], \ n_2\in[1.5, 4],  \  h\in[0.2p, 0.8p], \ w\in[0.2p, 0.8p]. 
\end{align}
Note that the line-shape of the resonance is the same after a simultaneous change of the signs of both $\alpha$ and $\tau$. Following our previous work \cite{Molokeev2023} we focus on the wavelengh range $1~100-2~000~{\rm nm}$. Thus, we specify the following ranges for the property parameters
\begin{align}\label{range2}
\tau \in [-1, 1], \ g \in [0, 1], \
\ 2~000~{\rm nm}> \lambda > 1~100~{\rm nm}, \ \gamma >0, 
\end{align}
where $\lambda$ is the wavelength of the resonance $\lambda=2\pi c/\omega_0$ and
\begin{equation}
    g=\sqrt{1-\tau^2}\sin(2\alpha).
\end{equation}
\begin{figure}[t!]
 \centering
\includegraphics[width=135mm]{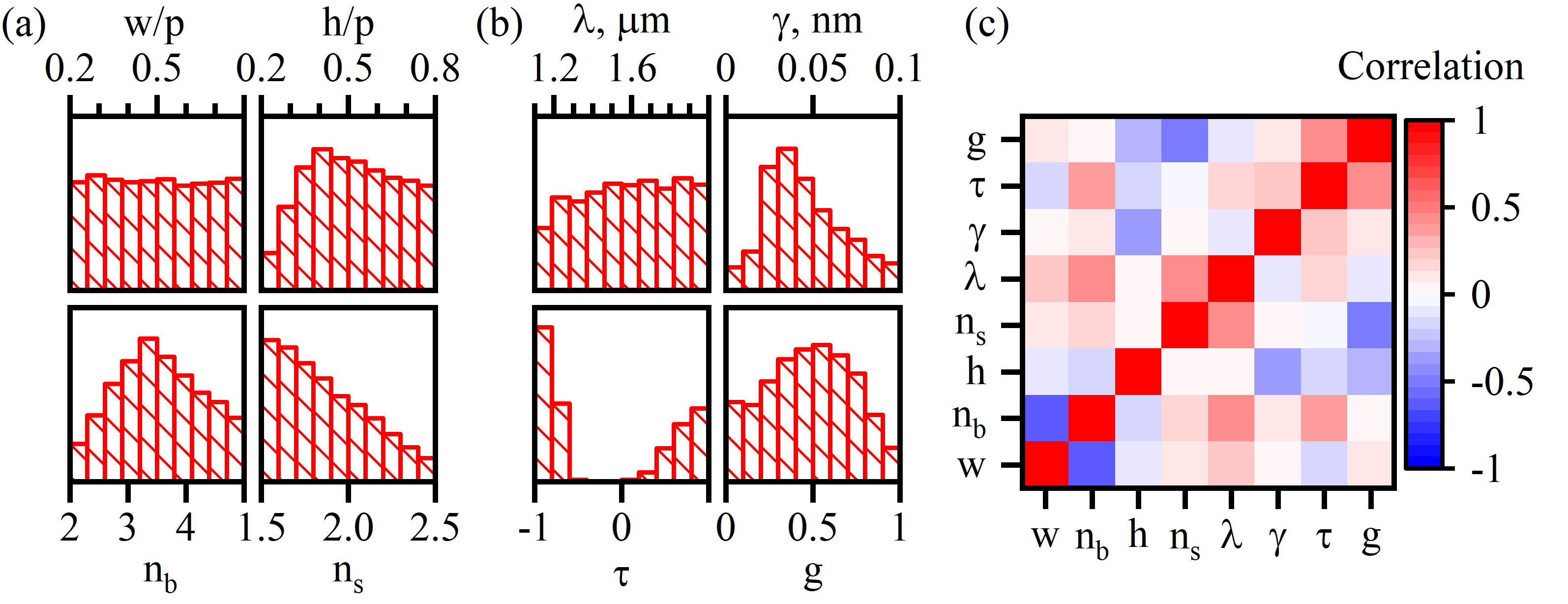}
\caption{Distribution of (a) -- feature parameters and (b) -- calculated  values of preperty parameters. (c) --  Correlation matrix between the feature and the property parameters.
} \label{figA1}
\end{figure}

To produce the data set we ran $100~000$ numerical experiments of which $18~836$ resulted in finding a symmetry protected BIC in the frequency range of interest. The simulations were preformed with application of the finite-element method (FEM) in COMSOL multiphysics package. The calculated values of the property parameters were extracted by least square fitting of Eq.~\eqref{TT} to the numerical data. The numerical experiments yielded values of four feature parameters ($h$, $n_b$, $w$, $n_s$) and four property parameters ($\lambda$, $g$, $\tau$, $\gamma$). In Fig.~\ref{figA1}~(a) we show the distribution of the feature parameters whereas the distribution of the property parameters is shown in Fig.~\ref{figA1}~(b). One can see in Figs.~\ref{figA1}~(a,~b)
that both the feature parameters and the BIC wavelength (property) exhibited almost uniform distributions. These uniform distributions suggest that the dataset encompasses representative cases. Additionally, the correlation matrix shown in Fig.~\ref{figA1}~(c) demonstrates the absence of linear relationships between the feature and property parameters, thereby justifying the application of machine learning methods.

\section{Random Forest method}
\label{sec:3}
In order to improve the precision of predicting the property parameters ($\tau$, $g$, $\lambda$, $\gamma$), we extended the dataset of feature parameters ($h$, $n_b$, $w$, $n_s$). incorporating their multiplication products, so the extended feature dataset also includes ($h^2$, $h\cdot n_b$, $h \cdot w$, $h\cdot n_s$, $n_b^2$, $n_b \cdot w$, $n_b \cdot n_s$, $w^2$, $w \cdot n_s$, $n_s^2$).
For the prediction task, we utilized the random forest (RF) algorithm, a powerful ensemble method based on regression trees~\cite{Breiman2001,Ho1995}. This approach involved constructing multiple decision trees by recursively partitioning the multidimensional predictor space. During the prediction phase, the RF model outputs the mode of the classes (for classification) or the mean average prediction (for regression) derived from the individual trees \cite{Liu2012,Segal1993}.
For the implementation of the RF model, we developed a Python script named RandomForest.py using the Python~3.6 programming language~\cite{Van_Rossum2016-yc}. The script utilises the standard libraries, including {\it numpy, pandas, sklearn, matplotlib,} and {\it mpl\_toolkits}. To account for the stochastic nature of the RF algorithm, we performed the 5-fold  cross-validation test, aggregating the results to obtain an averaged performance and calculate the mean average error (MAE). Each iteration involved randomly splitting the data into two sets. One set comprising 70\% of the total data was used for training the model. The remaining 30\% of the data was used for testing. As a result, we constructed four distinct RF models, one for each property parameter ($\tau$, $g$, $\lambda$, $\gamma$).  

Besides it predictive power, the RF algorithm can also quantify the importance of each feature parameter after training. This can be achieved by permuting the values of a selected feature within the training data and calculating the error on the perturbed dataset.  The importance score for the feature is obtained by averaging the difference in error before and after permutation across all trees and subsequent normalization by the differences \cite{Altmann2010,Wehenkel2018}. The features that yield higher values of  this score are ranked as more important compared to features with lower values.

\begin{figure}[t]
    \centering
    \includegraphics[width=135mm]{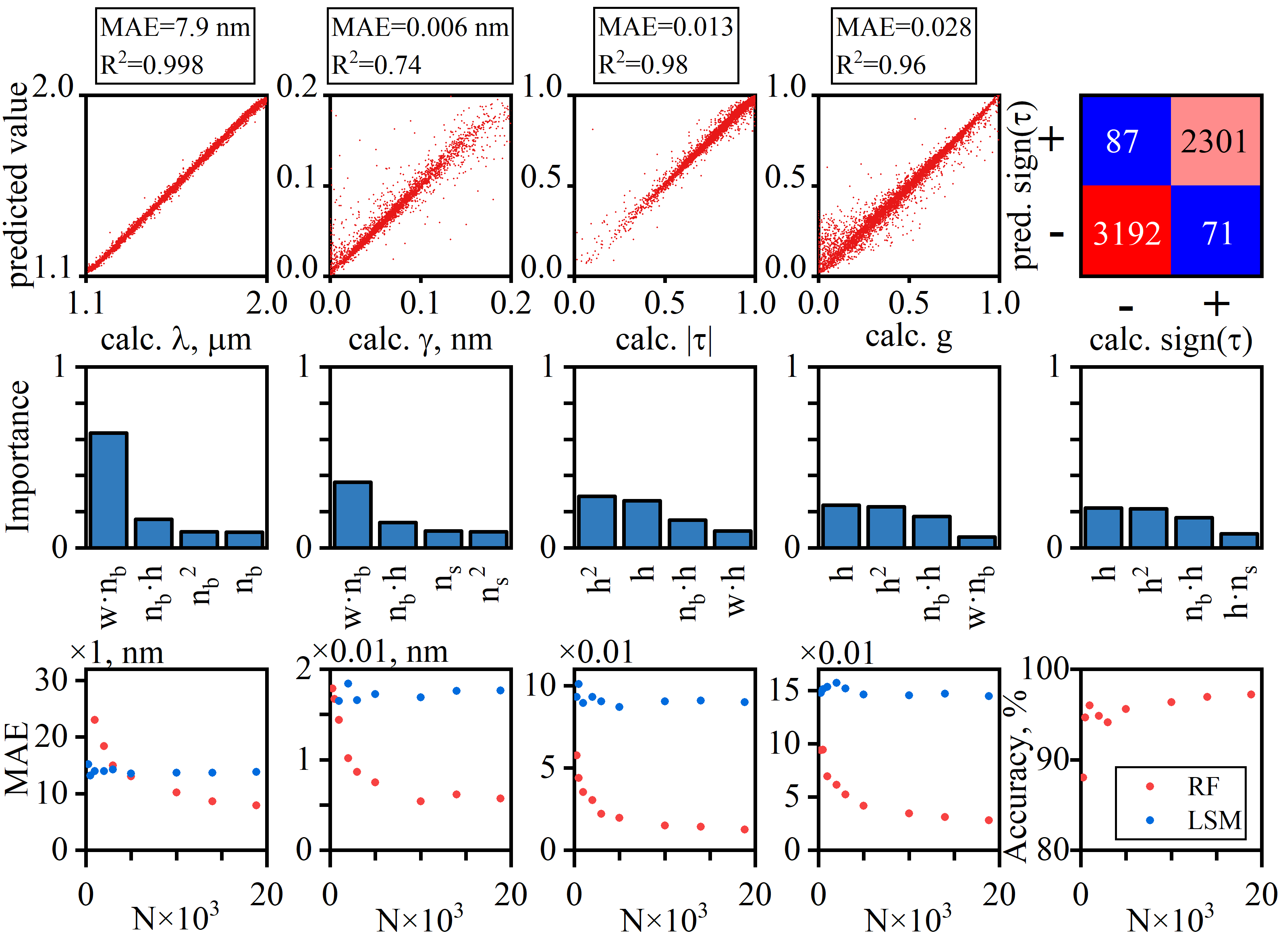}
     \caption{RF predictions for the four continuous property parameters $(\lambda, \gamma, |\tau|, g)$, and binary parameter ${\rm sign}(\tau)$. The first row shows the calculated vs. predicted values of the property parameters. Not that for the binary parameters  ${\rm sign}(\tau)$ the results are visualized in the form of the confusing matrix. The second row shows the histogram plots of the four larges importance scores for the extended feature parameter set. The third row compares the performance of the RF method against the LSM depending on the size of the training dataset $N$.}
    \label{fig:res_all}
\end{figure}

Finally, in applying the RF method, it is found out that the algorithm fails to correctly predict the property parameter $\tau$ when its absolute value approaches unity. This is due to the structure of Eq.~\eqref{TT} in which the numerator becomes independent of the sign of $\tau$ when $\rho/\tau\ll 1$. To amend this difficulty we used (0,1) binary representation of ${\rm sign}(\tau)$ to be predicted using the classification RF method. The quantity $|\tau|$ was used as the property parameter in application of the prediction RF instead of $\tau$. The results of application of the RF algorithm are collected in Fig.~\ref{fig:res_all}. In in Fig.~\ref{fig:res_all} we plot the RF predicted versus calculated values of the four continuous property parameters ($|\tau|$, $g$, $\lambda$, $\gamma$). The plots are supplemented by histograms of the importance score of four the most important feature parameters, and by plots comparing the RF performance against the polynomial least square method (LSM). In the case of the binary parameter the performance is qualified by the confusion matrix.


\section{Results}
\label{sec:4}
We proceed to anylizing the data presented in Fig.~\ref{fig:res_all}. In the first row of Fig.~\ref{fig:res_all} we plot the calculated versus the RF predicted values of the four continuous properties from the test data set. The last plot in the first row is the confusion matrix for ${\rm sign}(\tau)$. The MAE and the coefficients of determination $R^2$ for all continuous properties are presented on the top of each plot. One can see that all properties are predicted to a good accuracy with exception of $\gamma$. This is due to the singular behaviour of the resonant linewidth in the spectral vicinity of a BIC. Namely, since $\gamma$ can be vanishingly small its value can change by orders in magnitude under small variation of angle $\theta$ near normal incidence, see Eq.~\eqref{Taylor1}.  In the second row of Fig.~\ref{fig:res_all} we plot the importance score of the four most important features from the extended data set in predicting all five property parameters. The data show that with the exception of the 
wavelength, which is predominantly determined by the optical path across the bar \cite{Molokeev2023}, no other property is solely determined by a single parameter from the extended feature parameter set. In the third row of Fig.~\ref{fig:res_all} we compare the performance of RF against that of the  polynomial LSM approximation in dependence on the size of the training data set. One can see that for all properties except the resonant wavelength, the RF significantly outperforms the LSM. Moreover, for the properties $|\tau|$ and $g$, which solely determine the resonant line-shape the MAE of the RF is $\approx$ one order of magnitude smaller that that of the LSM. Note that comparison against the least square method is impossible for the binary property ${\rm {sign}}(\tau)$, and therefor no LSM data are presented in the last plot of the third row in Fig.~\ref{fig:res_all}.

\begin{figure}[t]
 \centering
\includegraphics[ width=135mm]{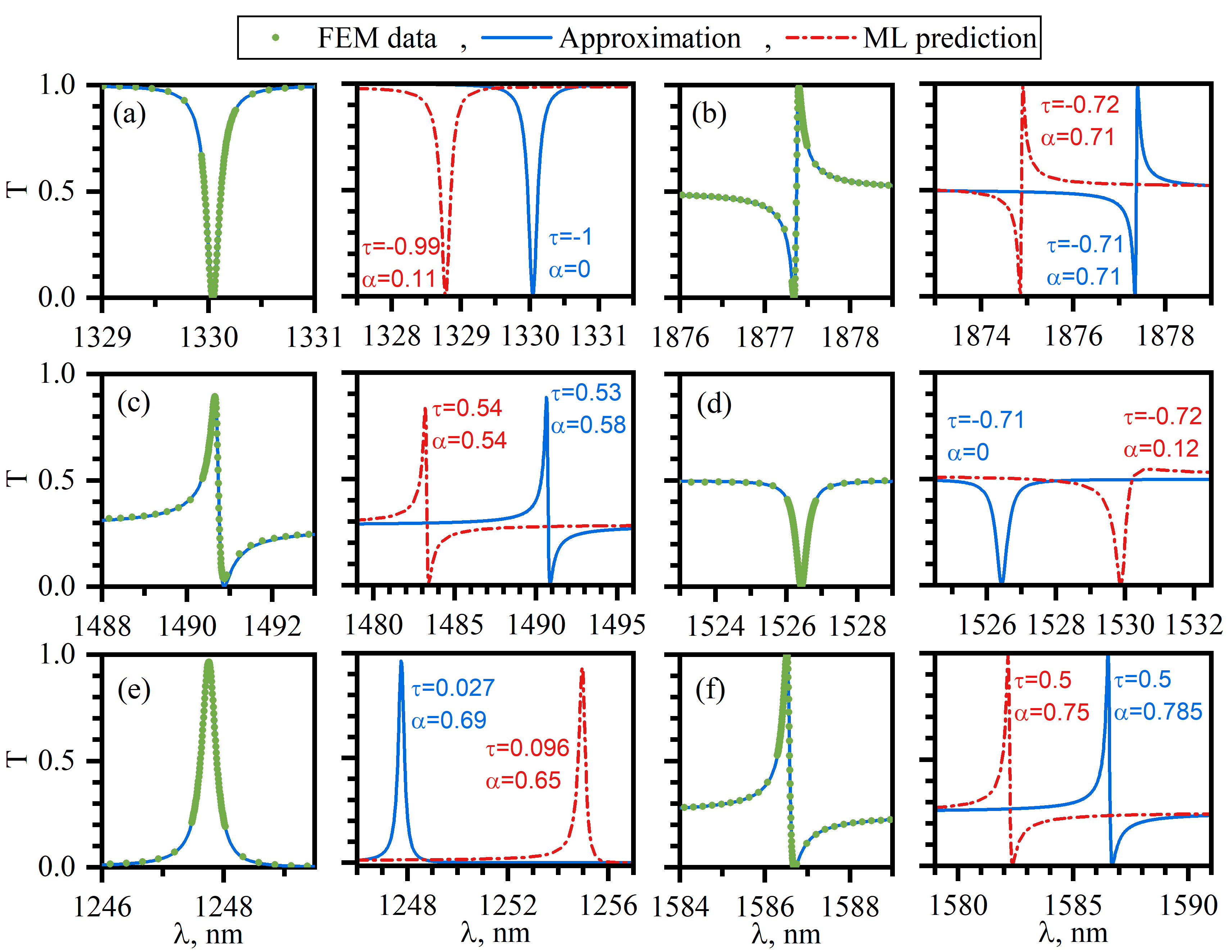}
\caption{Fano resonances induced by symmetry protected BICs in dielectric gratings. FEM calculation data -- green circles, Data approximation by Eq.~\eqref{TT} -- solid blue lines, RF -- predictions within the test dataset -- dash-dot red lines. The numerically obtained and the RF predicted values of $\alpha$ and $\tau$ are shown in each supblot.}
\label{ML_approx}
\end{figure}

The data collected allows one to draw some conclusion on the shapes of Fano resonance induced by symmetry protected BIC. First of all the position of the resonance is dictated by wavelengh $\lambda$ corresponding to the resonant frequency $\omega_0$. This quantity can be accurately predicted by both the RF method and the LSM due to the fact that is is predominantly controlled by a single feature $w\cdot n_b$. The width of the resonance, albeit it is better predicted by the RF method than by the LSM, is the most difficult to predict due to the singular nature of the BIC. This does not, however, impose a difficulty since in any prefabricated set-up the resonant width is easily controlled by the angle of incidence. Finally, the line-shape of the resonance is controlled by $\tau$ and $\alpha$, both being efficiently predicted by the RF method. It is worth mentioning that according to the data from Fig.~\ref{figA1} the distribution of $\tau$ has the following momenta $\langle |\tau| \rangle=0.791$, and $\langle \tau^2 \rangle=0.653$. It means that more often than not the resonance is observed on the background with transmission dominating over reflection. The first two momenta of the distribution of $\alpha$ are as follows $\langle \alpha \rangle=0.577$,  and $\langle \alpha^2 \rangle=0.376$. Remarkably, on average $\alpha$ is close to $\pi/4=0.785$ which corresponds to metasurfaces with up-down mirror symmetry. As it has been already mentioned the unit transmittance only occurs at $\alpha=\pi/4$. Thus, statistically the observed Fano resonances are likely to exhibit near-unit transmittance at the peak of Fano resonances.

The profiles of Fano resonances are demonstrated in Fig.~\eqref{ML_approx} where we plot six different line-shapes from the test dataset. For each case we present the FEM data are first compared against their approximation by Eq.~\eqref{TT} on the left of each subplot. On the right of each subplot we demonstrate the RF predicted line-shapes in comparison against Eq.~\eqref{TT}. One can see that in each case the position of the RF predicted resonance is shifted with respect to the calculated one by distance bigger the the line-width of the resonance. This  is due to vanishingly small line-width of the resonances with the average quality factor $Q = 8166$  in the spectral vicinity of a BIC across the . Note that, although the according to Fig.~\ref{fig:res_all} the resonant wavelength is predicted to a good accuracy, the RF fails to correctly position the resonance on the scale of its line-width. On the contrary the the RF predicted line-shapes of the resonance fit well to the calculated data. Note that different line-shapes are possible in the system under scrutiny including asymmetric Fano Fig.~\ref{ML_approx}~(b, c, f), Lorentzian Fig.~\ref{ML_approx}~(e), and anti-Lorentzian Fig.~\ref{ML_approx}~(a, d) line-shapes. Note that the transmittance always reaches zero at the dip of the resonances at the same time the numerically exact reflectance zeros are clearly absent in Fig.\ref{ML_approx}~(c, d).

\section{Conclusion}
\label{sec:concl}

In this work we investigated line-shapes of the Fano resonances induced by symmetry protected bound states in the continuum in dielectric gratings. It is numerically demonstrated that the line-shapes are controlled by two parameters in Eq.~\eqref{TT} which encompasses various line-shapes including asymmetric Fano, Lorentzian, and anti-Lorentzian resonances. In full accordance with the previous studies \cite{Gippius2005, Wang2013,  Yuan2022} it is confirmed that the transmittance zeros are always present even in the absence up-down symmetry. At the same time the reflectance zeros can only be approached accidentally when parameter $\alpha$ in Eq.~\eqref{TT} is close to $\pi/4$. It is found that the line-shapes of Fano resonances can be predicted to a good accuracy by the random forest machine learning methods which outperforms the standard least square methods approximation in error by an order of magnitude in error with the training dataset size $N\approx 10^4$. We speculate that the results presented can be of use in application for resonant response synthesis from all-delectric metasurfces.

\begin{acknowledgments}
This study was supported by the Ministry of  Science and Higher Education of Russian Federation (grant FSRZ-2023-0006), A.K. acknowledges the support by the Foundation for the Advancement of Theoretical Physics and Mathematics “BASIS” (grant 23-1-5-76-1), M.M. acknowledges the support by the Ministry of  Science and Higher Education of Russian Federation (grant FEWZ-2024-0052), V.G. acknowledges the support by the Ministry of  Science and Higher Education of Russian Federation (grant 124012900550-1).

\end{acknowledgments}

%

\appendix
\section{$S$-matrix reduction}\label{App1}
We start by rewriting Eq.~\eqref{S4} for the $4\times 4$ $S$-matrix
\begin{equation}\label{0A3}
\widehat{S}_{\scs{4}}=
\left(
\begin{array}{cc}
0 & \widehat{S}_{\scs{2}} \\
\widehat{S}^{\intercal}_{\scs{2}} & 0
\end{array}
\right)
\end{equation} 
The symmetry operation of the group $C_2^z$ has a matrix representation
\begin{equation}\label{0AP}
   \widehat{P} =
   \left(
   \begin{array}{cccc}
   0 & 0 & 1 & 0 \\
   0 & 0 & 0 & 1 \\
   1 & 0 & 0 & 0 \\
   0 & 1 & 0 & 0 
   \end{array}
   \right).
\end{equation}
in the space of the incident/outgoing amplitude vectors. Now assuming that the channel functions are defined to be symmetric with respect to the mirror operation Eq.~\eqref{0AP} one can state
\begin{equation}\label{0A4}
\widehat{S}_{\scs{4}}=\widehat{P}^{-1}\widehat{S}_{\scs{4}}\widehat{P}.
\end{equation}
After substituting Eq.~\eqref{0A3} to Eq.~\eqref{0A4} one finds
\begin{equation}
\widehat{S}_2=\widehat{S}_2^{\intercal}.
\end{equation}

\section{Derivation of Eq.~\eqref{TCMTcond}}\label{App2}

The first line in Eq.~\eqref{TCMTcond} can be proven by the same method as suggested in \cite{Fan03}. First of all, we assume that the resonant eigenmode is normalized to 
carry a unit energy whereas the scattering carry a unit a energy per unit of time across the interface between the far-field and the scattering domain. Thus, the absence of incidence wave the energy conservation leads to
\begin{equation}
\frac{dE}{dt}=\frac{d|a|^2}{dt}={\bf d}^{\dagger}{\bf d}|a|^2
\end{equation}
with $E$ being the energy stored in the resonant mode. Given that the solution of the first line in Eq.~\eqref{TCMT} is
\begin{equation}
a(t)=a_0e^{-(i\omega_0+\gamma) t}
\end{equation}
we immediately have
\begin{equation}
{\bf d}^{\dagger}{\bf d}=2\gamma.
\end{equation}

The derivation of the other relationships in Eq.~\eqref{TCMTcond} is more complicated. We start from the time-harmonic substitution in Eq.~\eqref{TCMT} which leads to the time-stationary TCMT equations in the following form
\begin{align}\label{0Asingle}
    & [i(\omega_0-\omega)+\gamma] { a}={\bf \kappa}^{\intercal}
    {\bf s}^{\smallp}\nonumber, \\
    & {\bf s}^{\smallm} = 
    \widehat{C}{\bf s}^{\smallp}+{\bf d} a.
\end{align}
The solution of Eq.\eqref{0Asingle} can be written in the form of $S$-matrix
\begin{equation}\label{0AS}
\widehat{S}_2(\omega)=\widehat{C}+\frac{{\bf d}{\bf \kappa}^{\intercal}}{i(\omega_0-\omega)+\gamma}.
\end{equation}
We notice that since $\widehat{S}_2(\infty)=\widehat{C}$ the matrix $\widehat{C}$ has the same symmetry and unitarity properties as $\widehat{S}_2(\omega)$. After applying $\widehat{S}_2^{-1}=\widehat{S}_2^{\dagger}$ we find
\begin{align}\label{0A6}
& \widehat{C}^*{\bf d}{\bf \kappa}^{\intercal}[-i(\omega_0-\omega)+\gamma]+
{\bf \kappa}^*{\bf d}^{\dagger}\widehat{C}[i(\omega_0-\omega)+\gamma] \nonumber \\
& +
2\gamma {\bf \kappa}^{*}{\bf \kappa}^{\intercal}=0.
\end{align}
By considering $\omega$-dependent terms one finds that
\begin{equation}
{\bf \kappa}^*{\bf d}^{\dagger}\widehat{C}=\widehat{C}^*{\bf d}{\bf \kappa}^{\intercal}.
\end{equation}
Applying the above equation in Eq.~\eqref{0A6} and examining the terms independent of $\omega$ we write
\begin{equation}
\widehat{C}^*{\bf d}{\bf \kappa}^{\intercal}= -{\bf \kappa}^{*}{\bf \kappa}^{\intercal}
\end{equation}
Next, assuming that the coupling vector ${\bf \kappa}$ has at least one non-zero element we have
\begin{equation}\label{0A7}
\widehat{C}{\bf d}^*= -{\bf \kappa}.
\end{equation}

Let us now utilize the symmetry of the scattering matrix. After applying $\widehat{S}_2=\widehat{S}_2^{\intercal}$ in Eq.~\eqref{0AS} we immediately have
\begin{align}\label{0A8}
{\bf d}{\bf \kappa}^{\intercal}={\bf \kappa}{\bf d}^{\intercal}.
\end{align}
Multiplying the above equation from the left by ${\bf d}^{\dagger}$ one obtains
\begin{equation}\label{0A9}
{\bf \kappa}=\frac{{\bf d}^{\dagger}{\bf \kappa}}{2\gamma}{\bf d}.
\end{equation}
Alternatively, by multiplying Eq.~\eqref{0A8} from the left by ${\bf \kappa}^{\dagger}$ one has
\begin{equation}
{\bf d}=\frac{{\bf \kappa}^{\dagger}{\bf d}}{{\bf \kappa}^{\dagger}{\bf \kappa}}{\bf \kappa}.
\end{equation}
The latter two equations combined result in
\begin{equation}\label{0A10}
2\gamma {\kappa}^{\dagger}{\bf \kappa}= |{\bf d}^{\dagger}{\bf \kappa}|^2
\end{equation}
Next, by multiplying Eq.~\eqref{0A7} from the left by its Hermitian adjoint one obtains
\begin{equation}\label{0A11}
2\gamma= {\kappa}^{\dagger}{\bf \kappa}
\end{equation}
After analysing Eq.~\eqref{0A9}, Eq.~\eqref{0A10}, and Eq.~\eqref{0A11} one can write
\begin{equation}\label{0A12}
{\bf d}=e^{i\eta}{\bf \kappa}, \ \eta\in[0, 2\pi).
\end{equation}

Let us now give interpretation to the phase $\eta$. By recollecting that the resonant mode is normalized to carry a unit energy we immediately see that its normalization constant is defined up to an arbitrary phase factor. Thus, all the TCMT equations have to be invariant under the $\rm{U}(1)$ transformation
\begin{equation}
a=e^{i\alpha}a'
\end{equation}
By plugging the above into Eq.~\eqref{0Asingle} we find that it remains invariant, i.e.~the same for the primed quantities, if the coupling and decoupling vectors are transformed as follows
\begin{align}
& {\bf d}=e^{-i\alpha}{\bf d}', \nonumber \\
& {\bf \kappa}=e^{+i\alpha}{\bf \kappa}'.
\end{align}
By choosing $\alpha=\eta/2$ one derives from Eq.~\eqref{0A7} and Eq.~\eqref{0A12}
\begin{align}
& \widehat{C}({\bf d}')^{*}+{\bf d}'=0, \nonumber \\
& {\bf \kappa}'={\bf d}'.
\end{align}
This equations are identical to the last two lines in Eq.~\eqref{TCMTcond}.

\section{Decoupling vector}\label{App3}
We start with the third line in Eq.~\eqref{TCMTcond} that reads \begin{equation}\label{A1}
\widehat{C}{\bf d}^*=-{\bf d}.
\end{equation}
Vector ${\bf d}$ is in general parameterized by four independent real numbers
\begin{equation}
    {\bf d}=
    \left(
    \begin{array}{c}
    a_{\scs{\rm (r)}}+ia_{\scs{\rm (i)}} \\
    b_{\scs{\rm (r)}}+ib_{\scs{\rm (i)}} 
    \end{array}
    \right).
\end{equation}
Substituting the above into Eq.~\eqref{A1} one obtains a set of linear homogeneous equations of rank two
\begin{equation}
\left(
\begin{array}{cccc}
     1+\rho & 0 & 0 & \tau \\
     0 & 1-\rho & \tau & 0 \\
     0 & \tau & 1+\rho & 0 \\
     \tau & 0 & 0 & 1-\rho
\end{array}
\right)
\left(
\begin{array}{c}
a_{\rm \scs{(r)}}\\
a_{\rm \scs{(i)}}\\
b_{\rm \scs{(r)}}\\
b_{\rm \scs{(i)}}
\end{array}
\right)=0,
\end{equation}
which has the general two-parametric solution
\begin{equation}\label{A2}
    {\bf d}=
    \left(
    \begin{array}{c}
        \tau b_{\rm\scs{(i)}}-i(1+\rho) a_{\rm\scs{(i)}}   \\
        \tau a_{\rm\scs{(i)}}-i(1+\rho) b_{\rm\scs{(i)}} 
    \end{array}
    \right).
\end{equation}
By recollecting that according to Eq.~\eqref{TCMTcond} $2\gamma=\bf{d}^{\dagger}\bf{d}$
one can find
\begin{equation}\label{A4}
    \gamma=(1+\rho)(a_{\rm \scs{(i)}}^2+b_{\rm \scs{(i)}}^2).
\end{equation}
After combining Eq.~\eqref{A2} with Eq.~\eqref{A4} one arrives at the general solution of the form
\begin{equation}
    {\bf d}\!=\!\sqrt{\frac{\gamma}{(1\!+\!\rho)}}
    \!\left(\!
    \begin{array}{c}
    \!\tau\cos\alpha-i(1\!+\!\rho) \sin\alpha   \\
        \!\tau \sin\alpha-i(1\!+\!\rho) \cos\alpha 
    \end{array}
    \!\right)\!, \ \alpha\!\in\!\left[-{\pi}/{2}, {\pi}/{2}\right],
\end{equation}
where the range $[-{\pi}/{2}, {\pi}/{2}]$ is chosen since ${\bf \kappa}={\bf d}$ and, therefore, the sign of $\bf{d}$ is not important for the $S$-matrix Eq.~\eqref{0AS}. 


\begin{thebibliography}{71}%
\makeatletter
\providecommand \@ifxundefined [1]{%
 \@ifx{#1\undefined}
}%
\providecommand \@ifnum [1]{%
 \ifnum #1\expandafter \@firstoftwo
 \else \expandafter \@secondoftwo
 \fi
}%
\providecommand \@ifx [1]{%
 \ifx #1\expandafter \@firstoftwo
 \else \expandafter \@secondoftwo
 \fi
}%
\providecommand \natexlab [1]{#1}%
\providecommand \enquote  [1]{``#1''}%
\providecommand \bibnamefont  [1]{#1}%
\providecommand \bibfnamefont [1]{#1}%
\providecommand \citenamefont [1]{#1}%
\providecommand \href@noop [0]{\@secondoftwo}%
\providecommand \href [0]{\begingroup \@sanitize@url \@href}%
\providecommand \@href[1]{\@@startlink{#1}\@@href}%
\providecommand \@@href[1]{\endgroup#1\@@endlink}%
\providecommand \@sanitize@url [0]{\catcode `\\12\catcode `\$12\catcode
  `\&12\catcode `\#12\catcode `\^12\catcode `\_12\catcode `\%12\relax}%
\providecommand \@@startlink[1]{}%
\providecommand \@@endlink[0]{}%
\providecommand \url  [0]{\begingroup\@sanitize@url \@url }%
\providecommand \@url [1]{\endgroup\@href {#1}{\urlprefix }}%
\providecommand \urlprefix  [0]{URL }%
\providecommand \Eprint [0]{\href }%
\providecommand \doibase [0]{https://doi.org/}%
\providecommand \selectlanguage [0]{\@gobble}%
\providecommand \bibinfo  [0]{\@secondoftwo}%
\providecommand \bibfield  [0]{\@secondoftwo}%
\providecommand \translation [1]{[#1]}%
\providecommand \BibitemOpen [0]{}%
\providecommand \bibitemStop [0]{}%
\providecommand \bibitemNoStop [0]{.\EOS\space}%
\providecommand \EOS [0]{\spacefactor3000\relax}%
\providecommand \BibitemShut  [1]{\csname bibitem#1\endcsname}%
\let\auto@bib@innerbib\@empty
\bibitem [{\citenamefont {Hsu}\ \emph {et~al.}(2016)\citenamefont {Hsu},
  \citenamefont {Zhen}, \citenamefont {Stone}, \citenamefont {Joannopoulos},\
  and\ \citenamefont {Solja{\v{c}}i{\'c}}}]{Hsu16}%
  \BibitemOpen
  \bibfield  {author} {\bibinfo {author} {\bibfnamefont {C.~W.}\ \bibnamefont
  {Hsu}}, \bibinfo {author} {\bibfnamefont {B.}~\bibnamefont {Zhen}}, \bibinfo
  {author} {\bibfnamefont {A.~D.}\ \bibnamefont {Stone}}, \bibinfo {author}
  {\bibfnamefont {J.~D.}\ \bibnamefont {Joannopoulos}},\ and\ \bibinfo {author}
  {\bibfnamefont {M.}~\bibnamefont {Solja{\v{c}}i{\'c}}},\ }\bibfield  {title}
  {\bibinfo {title} {Bound states in the continuum},\ }\href
  {https://doi.org/10.1038/natrevmats.2016.48} {\bibfield  {journal} {\bibinfo
  {journal} {Nature Reviews Materials}\ }\textbf {\bibinfo {volume} {1}},\
  \bibinfo {pages} {16048} (\bibinfo {year} {2016})}\BibitemShut {NoStop}%
\bibitem [{\citenamefont {Koshelev}\ \emph
  {et~al.}(2019{\natexlab{a}})\citenamefont {Koshelev}, \citenamefont
  {Bogdanov},\ and\ \citenamefont {Kivshar}}]{koshelev2019meta}%
  \BibitemOpen
  \bibfield  {author} {\bibinfo {author} {\bibfnamefont {K.}~\bibnamefont
  {Koshelev}}, \bibinfo {author} {\bibfnamefont {A.}~\bibnamefont {Bogdanov}},\
  and\ \bibinfo {author} {\bibfnamefont {Y.}~\bibnamefont {Kivshar}},\
  }\bibfield  {title} {\bibinfo {title} {Meta-optics and bound states in the
  continuum},\ }\href@noop {} {\bibfield  {journal} {\bibinfo  {journal}
  {Science Bulletin}\ }\textbf {\bibinfo {volume} {64}},\ \bibinfo {pages}
  {836} (\bibinfo {year} {2019}{\natexlab{a}})}\BibitemShut {NoStop}%
\bibitem [{\citenamefont {Koshelev}\ \emph
  {et~al.}(2019{\natexlab{b}})\citenamefont {Koshelev}, \citenamefont
  {Favraud}, \citenamefont {Bogdanov}, \citenamefont {Kivshar},\ and\
  \citenamefont {Fratalocchi}}]{Koshelev19}%
  \BibitemOpen
  \bibfield  {author} {\bibinfo {author} {\bibfnamefont {K.}~\bibnamefont
  {Koshelev}}, \bibinfo {author} {\bibfnamefont {G.}~\bibnamefont {Favraud}},
  \bibinfo {author} {\bibfnamefont {A.}~\bibnamefont {Bogdanov}}, \bibinfo
  {author} {\bibfnamefont {Y.}~\bibnamefont {Kivshar}},\ and\ \bibinfo {author}
  {\bibfnamefont {A.}~\bibnamefont {Fratalocchi}},\ }\bibfield  {title}
  {\bibinfo {title} {Nonradiating photonics with resonant dielectric
  nanostructures},\ }\href {https://doi.org/10.1515/nanoph-2019-0024}
  {\bibfield  {journal} {\bibinfo  {journal} {Nanophotonics}\ }\textbf
  {\bibinfo {volume} {8}},\ \bibinfo {pages} {725} (\bibinfo {year}
  {2019}{\natexlab{b}})}\BibitemShut {NoStop}%
\bibitem [{\citenamefont {Joseph}\ \emph {et~al.}(2021)\citenamefont {Joseph},
  \citenamefont {Pandey}, \citenamefont {Sarkar},\ and\ \citenamefont
  {Joseph}}]{joseph2021}%
  \BibitemOpen
  \bibfield  {author} {\bibinfo {author} {\bibfnamefont {S.}~\bibnamefont
  {Joseph}}, \bibinfo {author} {\bibfnamefont {S.}~\bibnamefont {Pandey}},
  \bibinfo {author} {\bibfnamefont {S.}~\bibnamefont {Sarkar}},\ and\ \bibinfo
  {author} {\bibfnamefont {J.}~\bibnamefont {Joseph}},\ }\bibfield  {title}
  {\bibinfo {title} {Bound states in the continuum in resonant nanostructures:
  an overview of engineered materials for tailored applications},\ }\href@noop
  {} {\bibfield  {journal} {\bibinfo  {journal} {Nanophotonics}\ }\textbf
  {\bibinfo {volume} {10}},\ \bibinfo {pages} {4175} (\bibinfo {year}
  {2021})}\BibitemShut {NoStop}%
\bibitem [{\citenamefont {Kang}\ \emph {et~al.}(2023)\citenamefont {Kang},
  \citenamefont {Liu}, \citenamefont {Chan},\ and\ \citenamefont
  {Xiao}}]{Kang2023}%
  \BibitemOpen
  \bibfield  {author} {\bibinfo {author} {\bibfnamefont {M.}~\bibnamefont
  {Kang}}, \bibinfo {author} {\bibfnamefont {T.}~\bibnamefont {Liu}}, \bibinfo
  {author} {\bibfnamefont {C.~T.}\ \bibnamefont {Chan}},\ and\ \bibinfo
  {author} {\bibfnamefont {M.}~\bibnamefont {Xiao}},\ }\bibfield  {title}
  {\bibinfo {title} {Applications of bound states in the continuum in
  photonics},\ }\href {https://doi.org/10.1038/s42254-023-00642-8} {\bibfield
  {journal} {\bibinfo  {journal} {Nature Reviews Physics}\ }\textbf {\bibinfo
  {volume} {5}},\ \bibinfo {pages} {659} (\bibinfo {year} {2023})}\BibitemShut
  {NoStop}%
\bibitem [{\citenamefont {Zhang}\ and\ \citenamefont
  {Zhang}(2015)}]{zhang2015ultrasensitive}%
  \BibitemOpen
  \bibfield  {author} {\bibinfo {author} {\bibfnamefont {M.}~\bibnamefont
  {Zhang}}\ and\ \bibinfo {author} {\bibfnamefont {X.}~\bibnamefont {Zhang}},\
  }\bibfield  {title} {\bibinfo {title} {Ultrasensitive optical absorption in
  graphene based on bound states in the continuum},\ }\href@noop {} {\bibfield
  {journal} {\bibinfo  {journal} {Scientific reports}\ }\textbf {\bibinfo
  {volume} {5}},\ \bibinfo {pages} {1} (\bibinfo {year} {2015})}\BibitemShut
  {NoStop}%
\bibitem [{\citenamefont {Wang}\ \emph {et~al.}(2020)\citenamefont {Wang},
  \citenamefont {Duan}, \citenamefont {Chen}, \citenamefont {Zhou},
  \citenamefont {Liu},\ and\ \citenamefont {Xiao}}]{wang2020controlling}%
  \BibitemOpen
  \bibfield  {author} {\bibinfo {author} {\bibfnamefont {X.}~\bibnamefont
  {Wang}}, \bibinfo {author} {\bibfnamefont {J.}~\bibnamefont {Duan}}, \bibinfo
  {author} {\bibfnamefont {W.}~\bibnamefont {Chen}}, \bibinfo {author}
  {\bibfnamefont {C.}~\bibnamefont {Zhou}}, \bibinfo {author} {\bibfnamefont
  {T.}~\bibnamefont {Liu}},\ and\ \bibinfo {author} {\bibfnamefont
  {S.}~\bibnamefont {Xiao}},\ }\bibfield  {title} {\bibinfo {title}
  {Controlling light absorption of graphene at critical coupling through
  magnetic dipole quasi-bound states in the continuum resonance},\ }\href@noop
  {} {\bibfield  {journal} {\bibinfo  {journal} {Physical Review B}\ }\textbf
  {\bibinfo {volume} {102}},\ \bibinfo {pages} {155432} (\bibinfo {year}
  {2020})}\BibitemShut {NoStop}%
\bibitem [{\citenamefont {Sang}\ \emph {et~al.}(2021)\citenamefont {Sang},
  \citenamefont {Dereshgi}, \citenamefont {Hadibrata}, \citenamefont
  {Tanriover},\ and\ \citenamefont {Aydin}}]{sang2021highly}%
  \BibitemOpen
  \bibfield  {author} {\bibinfo {author} {\bibfnamefont {T.}~\bibnamefont
  {Sang}}, \bibinfo {author} {\bibfnamefont {S.~A.}\ \bibnamefont {Dereshgi}},
  \bibinfo {author} {\bibfnamefont {W.}~\bibnamefont {Hadibrata}}, \bibinfo
  {author} {\bibfnamefont {I.}~\bibnamefont {Tanriover}},\ and\ \bibinfo
  {author} {\bibfnamefont {K.}~\bibnamefont {Aydin}},\ }\bibfield  {title}
  {\bibinfo {title} {Highly efficient light absorption of monolayer graphene by
  quasi-bound state in the continuum},\ }\href@noop {} {\bibfield  {journal}
  {\bibinfo  {journal} {Nanomaterials}\ }\textbf {\bibinfo {volume} {11}},\
  \bibinfo {pages} {484} (\bibinfo {year} {2021})}\BibitemShut {NoStop}%
\bibitem [{\citenamefont {Xiao}\ \emph {et~al.}(2021)\citenamefont {Xiao},
  \citenamefont {Wang}, \citenamefont {Duan}, \citenamefont {Liu},\ and\
  \citenamefont {Yu}}]{xiao2021engineering}%
  \BibitemOpen
  \bibfield  {author} {\bibinfo {author} {\bibfnamefont {S.}~\bibnamefont
  {Xiao}}, \bibinfo {author} {\bibfnamefont {X.}~\bibnamefont {Wang}}, \bibinfo
  {author} {\bibfnamefont {J.}~\bibnamefont {Duan}}, \bibinfo {author}
  {\bibfnamefont {T.}~\bibnamefont {Liu}},\ and\ \bibinfo {author}
  {\bibfnamefont {T.}~\bibnamefont {Yu}},\ }\bibfield  {title} {\bibinfo
  {title} {Engineering light absorption at critical coupling via bound states
  in the continuum},\ }\href@noop {} {\bibfield  {journal} {\bibinfo  {journal}
  {JOSA B}\ }\textbf {\bibinfo {volume} {38}},\ \bibinfo {pages} {1325}
  (\bibinfo {year} {2021})}\BibitemShut {NoStop}%
\bibitem [{\citenamefont {Cai}\ \emph {et~al.}(2022)\citenamefont {Cai},
  \citenamefont {Liu}, \citenamefont {Zhu}, \citenamefont {Wu},\ and\
  \citenamefont {Huang}}]{cai2022enhancing}%
  \BibitemOpen
  \bibfield  {author} {\bibinfo {author} {\bibfnamefont {Y.}~\bibnamefont
  {Cai}}, \bibinfo {author} {\bibfnamefont {X.}~\bibnamefont {Liu}}, \bibinfo
  {author} {\bibfnamefont {K.}~\bibnamefont {Zhu}}, \bibinfo {author}
  {\bibfnamefont {H.}~\bibnamefont {Wu}},\ and\ \bibinfo {author}
  {\bibfnamefont {Y.}~\bibnamefont {Huang}},\ }\bibfield  {title} {\bibinfo
  {title} {Enhancing light absorption of graphene with dual quasi bound states
  in the continuum resonances},\ }\href@noop {} {\bibfield  {journal} {\bibinfo
   {journal} {Journal of Quantitative Spectroscopy and Radiative Transfer}\
  }\textbf {\bibinfo {volume} {283}},\ \bibinfo {pages} {108150} (\bibinfo
  {year} {2022})}\BibitemShut {NoStop}%
\bibitem [{\citenamefont {Liu}\ \emph {et~al.}(2017)\citenamefont {Liu},
  \citenamefont {Zhou},\ and\ \citenamefont {Sun}}]{Liu17}%
  \BibitemOpen
  \bibfield  {author} {\bibinfo {author} {\bibfnamefont {Y.}~\bibnamefont
  {Liu}}, \bibinfo {author} {\bibfnamefont {W.}~\bibnamefont {Zhou}},\ and\
  \bibinfo {author} {\bibfnamefont {Y.}~\bibnamefont {Sun}},\ }\bibfield
  {title} {\bibinfo {title} {Optical refractive index sensing based on high-{Q}
  bound states in the continuum in free-space coupled photonic crystal slabs},\
  }\href {https://doi.org/10.3390/s17081861} {\bibfield  {journal} {\bibinfo
  {journal} {Sensors}\ }\textbf {\bibinfo {volume} {17}},\ \bibinfo {pages}
  {1861} (\bibinfo {year} {2017})}\BibitemShut {NoStop}%
\bibitem [{\citenamefont {Romano}\ \emph {et~al.}(2018)\citenamefont {Romano},
  \citenamefont {Zito}, \citenamefont {Torino}, \citenamefont {Calafiore},
  \citenamefont {Penzo}, \citenamefont {Coppola}, \citenamefont {Cabrini},
  \citenamefont {Rendina},\ and\ \citenamefont {Mocella}}]{Romano18b}%
  \BibitemOpen
  \bibfield  {author} {\bibinfo {author} {\bibfnamefont {S.}~\bibnamefont
  {Romano}}, \bibinfo {author} {\bibfnamefont {G.}~\bibnamefont {Zito}},
  \bibinfo {author} {\bibfnamefont {S.}~\bibnamefont {Torino}}, \bibinfo
  {author} {\bibfnamefont {G.}~\bibnamefont {Calafiore}}, \bibinfo {author}
  {\bibfnamefont {E.}~\bibnamefont {Penzo}}, \bibinfo {author} {\bibfnamefont
  {G.}~\bibnamefont {Coppola}}, \bibinfo {author} {\bibfnamefont
  {S.}~\bibnamefont {Cabrini}}, \bibinfo {author} {\bibfnamefont
  {I.}~\bibnamefont {Rendina}},\ and\ \bibinfo {author} {\bibfnamefont
  {V.}~\bibnamefont {Mocella}},\ }\bibfield  {title} {\bibinfo {title}
  {Label-free sensing of ultralow-weight molecules with all-dielectric
  metasurfaces supporting bound states in the continuum},\ }\href
  {https://doi.org/10.1364/prj.6.000726} {\bibfield  {journal} {\bibinfo
  {journal} {Photonics Research}\ }\textbf {\bibinfo {volume} {6}},\ \bibinfo
  {pages} {726} (\bibinfo {year} {2018})}\BibitemShut {NoStop}%
\bibitem [{\citenamefont {Ndangali}\ and\ \citenamefont
  {Shabanov}(2013)}]{ndangali2013resonant}%
  \BibitemOpen
  \bibfield  {author} {\bibinfo {author} {\bibfnamefont {F.~R.}\ \bibnamefont
  {Ndangali}}\ and\ \bibinfo {author} {\bibfnamefont {S.~V.}\ \bibnamefont
  {Shabanov}},\ }\bibfield  {title} {\bibinfo {title} {The resonant nonlinear
  scattering theory with bound states in the radiation continuum and the second
  harmonic generation},\ }in\ \href@noop {} {\emph {\bibinfo {booktitle}
  {Active Photonic Materials V}}},\ Vol.\ \bibinfo {volume} {8808}\ (\bibinfo
  {organization} {International Society for Optics and Photonics},\ \bibinfo
  {year} {2013})\ p.\ \bibinfo {pages} {88081F}\BibitemShut {NoStop}%
\bibitem [{\citenamefont {Wang}\ and\ \citenamefont
  {Zhang}(2018)}]{wang2018large}%
  \BibitemOpen
  \bibfield  {author} {\bibinfo {author} {\bibfnamefont {T.}~\bibnamefont
  {Wang}}\ and\ \bibinfo {author} {\bibfnamefont {S.}~\bibnamefont {Zhang}},\
  }\bibfield  {title} {\bibinfo {title} {Large enhancement of second harmonic
  generation from transition-metal dichalcogenide monolayer on grating near
  bound states in the continuum},\ }\href@noop {} {\bibfield  {journal}
  {\bibinfo  {journal} {Optics express}\ }\textbf {\bibinfo {volume} {26}},\
  \bibinfo {pages} {322} (\bibinfo {year} {2018})}\BibitemShut {NoStop}%
\bibitem [{\citenamefont {Carletti}\ \emph {et~al.}(2018)\citenamefont
  {Carletti}, \citenamefont {Koshelev}, \citenamefont {De~Angelis},\ and\
  \citenamefont {Kivshar}}]{carletti2018giant}%
  \BibitemOpen
  \bibfield  {author} {\bibinfo {author} {\bibfnamefont {L.}~\bibnamefont
  {Carletti}}, \bibinfo {author} {\bibfnamefont {K.}~\bibnamefont {Koshelev}},
  \bibinfo {author} {\bibfnamefont {C.}~\bibnamefont {De~Angelis}},\ and\
  \bibinfo {author} {\bibfnamefont {Y.}~\bibnamefont {Kivshar}},\ }\bibfield
  {title} {\bibinfo {title} {Giant nonlinear response at the nanoscale driven
  by bound states in the continuum},\ }\href@noop {} {\bibfield  {journal}
  {\bibinfo  {journal} {Physical review letters}\ }\textbf {\bibinfo {volume}
  {121}},\ \bibinfo {pages} {033903} (\bibinfo {year} {2018})}\BibitemShut
  {NoStop}%
\bibitem [{\citenamefont {Koshelev}\ \emph {et~al.}(2020)\citenamefont
  {Koshelev}, \citenamefont {Kruk}, \citenamefont {Melik-Gaykazyan},
  \citenamefont {Choi}, \citenamefont {Bogdanov}, \citenamefont {Park},\ and\
  \citenamefont {Kivshar}}]{koshelev2020subwavelength}%
  \BibitemOpen
  \bibfield  {author} {\bibinfo {author} {\bibfnamefont {K.}~\bibnamefont
  {Koshelev}}, \bibinfo {author} {\bibfnamefont {S.}~\bibnamefont {Kruk}},
  \bibinfo {author} {\bibfnamefont {E.}~\bibnamefont {Melik-Gaykazyan}},
  \bibinfo {author} {\bibfnamefont {J.-H.}\ \bibnamefont {Choi}}, \bibinfo
  {author} {\bibfnamefont {A.}~\bibnamefont {Bogdanov}}, \bibinfo {author}
  {\bibfnamefont {H.-G.}\ \bibnamefont {Park}},\ and\ \bibinfo {author}
  {\bibfnamefont {Y.}~\bibnamefont {Kivshar}},\ }\bibfield  {title} {\bibinfo
  {title} {Subwavelength dielectric resonators for nonlinear nanophotonics},\
  }\href@noop {} {\bibfield  {journal} {\bibinfo  {journal} {Science}\ }\textbf
  {\bibinfo {volume} {367}},\ \bibinfo {pages} {288} (\bibinfo {year}
  {2020})}\BibitemShut {NoStop}%
\bibitem [{\citenamefont {Kodigala}\ \emph {et~al.}(2017)\citenamefont
  {Kodigala}, \citenamefont {Lepetit}, \citenamefont {Gu}, \citenamefont
  {Bahari}, \citenamefont {Fainman},\ and\ \citenamefont
  {Kant\'{e}}}]{Kodigala17}%
  \BibitemOpen
  \bibfield  {author} {\bibinfo {author} {\bibfnamefont {A.}~\bibnamefont
  {Kodigala}}, \bibinfo {author} {\bibfnamefont {T.}~\bibnamefont {Lepetit}},
  \bibinfo {author} {\bibfnamefont {Q.}~\bibnamefont {Gu}}, \bibinfo {author}
  {\bibfnamefont {B.}~\bibnamefont {Bahari}}, \bibinfo {author} {\bibfnamefont
  {Y.}~\bibnamefont {Fainman}},\ and\ \bibinfo {author} {\bibfnamefont
  {B.}~\bibnamefont {Kant\'{e}}},\ }\bibfield  {title} {\bibinfo {title}
  {Lasing action from photonic bound states in continuum},\ }\href
  {https://doi.org/10.1038/nature20799} {\bibfield  {journal} {\bibinfo
  {journal} {Nature}\ }\textbf {\bibinfo {volume} {541}},\ \bibinfo {pages}
  {196} (\bibinfo {year} {2017})}\BibitemShut {NoStop}%
\bibitem [{\citenamefont {Hwang}\ \emph {et~al.}(2021)\citenamefont {Hwang},
  \citenamefont {Lee}, \citenamefont {Kim}, \citenamefont {Jeong},
  \citenamefont {Kwon}, \citenamefont {Koshelev}, \citenamefont {Kivshar},\
  and\ \citenamefont {Park}}]{hwang2021ultralow}%
  \BibitemOpen
  \bibfield  {author} {\bibinfo {author} {\bibfnamefont {M.-S.}\ \bibnamefont
  {Hwang}}, \bibinfo {author} {\bibfnamefont {H.-C.}\ \bibnamefont {Lee}},
  \bibinfo {author} {\bibfnamefont {K.-H.}\ \bibnamefont {Kim}}, \bibinfo
  {author} {\bibfnamefont {K.-Y.}\ \bibnamefont {Jeong}}, \bibinfo {author}
  {\bibfnamefont {S.-H.}\ \bibnamefont {Kwon}}, \bibinfo {author}
  {\bibfnamefont {K.}~\bibnamefont {Koshelev}}, \bibinfo {author}
  {\bibfnamefont {Y.}~\bibnamefont {Kivshar}},\ and\ \bibinfo {author}
  {\bibfnamefont {H.-G.}\ \bibnamefont {Park}},\ }\bibfield  {title} {\bibinfo
  {title} {Ultralow-threshold laser using super-bound states in the
  continuum},\ }\href {https://doi.org/10.1038/s41467-021-24502-0} {\bibfield
  {journal} {\bibinfo  {journal} {Nature Communications}\ }\textbf {\bibinfo
  {volume} {12}},\ \bibinfo {pages} {4135} (\bibinfo {year}
  {2021})}\BibitemShut {NoStop}%
\bibitem [{\citenamefont {Yu}\ \emph {et~al.}(2021)\citenamefont {Yu},
  \citenamefont {Sakanas}, \citenamefont {Zali}, \citenamefont {Semenova},
  \citenamefont {Yvind},\ and\ \citenamefont {M{\o}rk}}]{Yu2021ultra}%
  \BibitemOpen
  \bibfield  {author} {\bibinfo {author} {\bibfnamefont {Y.}~\bibnamefont
  {Yu}}, \bibinfo {author} {\bibfnamefont {A.}~\bibnamefont {Sakanas}},
  \bibinfo {author} {\bibfnamefont {A.~R.}\ \bibnamefont {Zali}}, \bibinfo
  {author} {\bibfnamefont {E.}~\bibnamefont {Semenova}}, \bibinfo {author}
  {\bibfnamefont {K.}~\bibnamefont {Yvind}},\ and\ \bibinfo {author}
  {\bibfnamefont {J.}~\bibnamefont {M{\o}rk}},\ }\bibfield  {title} {\bibinfo
  {title} {Ultra-coherent fano laser based on a bound state in the continuum},\
  }\href {https://doi.org/10.1038/s41566-021-00860-5} {\bibfield  {journal}
  {\bibinfo  {journal} {Nature Photonics}\ }\textbf {\bibinfo {volume} {15}},\
  \bibinfo {pages} {758} (\bibinfo {year} {2021})}\BibitemShut {NoStop}%
\bibitem [{\citenamefont {Yang}\ \emph {et~al.}(2021)\citenamefont {Yang},
  \citenamefont {Huang}, \citenamefont {Maksimov}, \citenamefont {Pankin},
  \citenamefont {Timofeev}, \citenamefont {Hong}, \citenamefont {Li},
  \citenamefont {Chen}, \citenamefont {Hsu}, \citenamefont {Liu}, \citenamefont
  {Lu}, \citenamefont {Lin}, \citenamefont {Yang},\ and\ \citenamefont
  {Chen}}]{yang2021low}%
  \BibitemOpen
  \bibfield  {author} {\bibinfo {author} {\bibfnamefont {J.-H.}\ \bibnamefont
  {Yang}}, \bibinfo {author} {\bibfnamefont {Z.-T.}\ \bibnamefont {Huang}},
  \bibinfo {author} {\bibfnamefont {D.~N.}\ \bibnamefont {Maksimov}}, \bibinfo
  {author} {\bibfnamefont {P.~S.}\ \bibnamefont {Pankin}}, \bibinfo {author}
  {\bibfnamefont {I.~V.}\ \bibnamefont {Timofeev}}, \bibinfo {author}
  {\bibfnamefont {K.-B.}\ \bibnamefont {Hong}}, \bibinfo {author}
  {\bibfnamefont {H.}~\bibnamefont {Li}}, \bibinfo {author} {\bibfnamefont
  {J.-W.}\ \bibnamefont {Chen}}, \bibinfo {author} {\bibfnamefont {C.-Y.}\
  \bibnamefont {Hsu}}, \bibinfo {author} {\bibfnamefont {Y.-Y.}\ \bibnamefont
  {Liu}}, \bibinfo {author} {\bibfnamefont {T.-C.}\ \bibnamefont {Lu}},
  \bibinfo {author} {\bibfnamefont {T.-R.}\ \bibnamefont {Lin}}, \bibinfo
  {author} {\bibfnamefont {C.-S.}\ \bibnamefont {Yang}},\ and\ \bibinfo
  {author} {\bibfnamefont {K.-P.}\ \bibnamefont {Chen}},\ }\bibfield  {title}
  {\bibinfo {title} {Low-threshold bound state in the continuum lasers in
  hybrid lattice resonance metasurfaces},\ }\href
  {https://doi.org/https://doi.org/10.1002/lpor.202100118} {\bibfield
  {journal} {\bibinfo  {journal} {Laser \& Photonics Reviews}\ }\textbf
  {\bibinfo {volume} {15}},\ \bibinfo {pages} {2100118} (\bibinfo {year}
  {2021})},\ \Eprint
  {https://arxiv.org/abs/https://onlinelibrary.wiley.com/doi/pdf/10.1002/lpor.202100118}
  {https://onlinelibrary.wiley.com/doi/pdf/10.1002/lpor.202100118} \BibitemShut
  {NoStop}%
\bibitem [{\citenamefont {Koshelev}\ \emph {et~al.}(2018)\citenamefont
  {Koshelev}, \citenamefont {Lepeshov}, \citenamefont {Liu}, \citenamefont
  {Bogdanov},\ and\ \citenamefont {Kivshar}}]{Koshelev18}%
  \BibitemOpen
  \bibfield  {author} {\bibinfo {author} {\bibfnamefont {K.}~\bibnamefont
  {Koshelev}}, \bibinfo {author} {\bibfnamefont {S.}~\bibnamefont {Lepeshov}},
  \bibinfo {author} {\bibfnamefont {M.}~\bibnamefont {Liu}}, \bibinfo {author}
  {\bibfnamefont {A.}~\bibnamefont {Bogdanov}},\ and\ \bibinfo {author}
  {\bibfnamefont {Y.}~\bibnamefont {Kivshar}},\ }\bibfield  {title} {\bibinfo
  {title} {Asymmetric metasurfaces with high-{Q} resonances governed by bound
  states in the continuum},\ }\bibfield  {journal} {\bibinfo  {journal}
  {Physical Review Letters}\ }\textbf {\bibinfo {volume} {121}},\ \href
  {https://doi.org/10.1103/physrevlett.121.193903}
  {10.1103/physrevlett.121.193903} (\bibinfo {year} {2018})\BibitemShut
  {NoStop}%
\bibitem [{\citenamefont {Maksimov}\ \emph
  {et~al.}(2020{\natexlab{a}})\citenamefont {Maksimov}, \citenamefont
  {Gerasimov}, \citenamefont {Romano},\ and\ \citenamefont
  {Polyutov}}]{Maksimov20}%
  \BibitemOpen
  \bibfield  {author} {\bibinfo {author} {\bibfnamefont {D.~N.}\ \bibnamefont
  {Maksimov}}, \bibinfo {author} {\bibfnamefont {V.~S.}\ \bibnamefont
  {Gerasimov}}, \bibinfo {author} {\bibfnamefont {S.}~\bibnamefont {Romano}},\
  and\ \bibinfo {author} {\bibfnamefont {S.~P.}\ \bibnamefont {Polyutov}},\
  }\bibfield  {title} {\bibinfo {title} {Refractive index sensing with optical
  bound states in the continuum},\ }\href {https://doi.org/10.1364/oe.411749}
  {\bibfield  {journal} {\bibinfo  {journal} {Optics Express}\ }\textbf
  {\bibinfo {volume} {28}},\ \bibinfo {pages} {38907} (\bibinfo {year}
  {2020}{\natexlab{a}})}\BibitemShut {NoStop}%
\bibitem [{\citenamefont {Shipman}\ and\ \citenamefont
  {Venakides}(2005)}]{Shipman}%
  \BibitemOpen
  \bibfield  {author} {\bibinfo {author} {\bibfnamefont {S.~P.}\ \bibnamefont
  {Shipman}}\ and\ \bibinfo {author} {\bibfnamefont {S.}~\bibnamefont
  {Venakides}},\ }\bibfield  {title} {\bibinfo {title} {Resonant transmission
  near nonrobust periodic slab modes},\ }\href@noop {} {\bibfield  {journal}
  {\bibinfo  {journal} {Physical Review E}\ }\textbf {\bibinfo {volume} {71}},\
  \bibinfo {pages} {026611} (\bibinfo {year} {2005})}\BibitemShut {NoStop}%
\bibitem [{\citenamefont {Sadreev}\ \emph {et~al.}(2006)\citenamefont
  {Sadreev}, \citenamefont {Bulgakov},\ and\ \citenamefont {Rotter}}]{SBR}%
  \BibitemOpen
  \bibfield  {author} {\bibinfo {author} {\bibfnamefont {A.~F.}\ \bibnamefont
  {Sadreev}}, \bibinfo {author} {\bibfnamefont {E.~N.}\ \bibnamefont
  {Bulgakov}},\ and\ \bibinfo {author} {\bibfnamefont {I.}~\bibnamefont
  {Rotter}},\ }\bibfield  {title} {\bibinfo {title} {Bound states in the
  continuum in open quantum billiards with a variable shape},\ }\href@noop {}
  {\bibfield  {journal} {\bibinfo  {journal} {Physical Review B}\ }\textbf
  {\bibinfo {volume} {73}},\ \bibinfo {pages} {235342} (\bibinfo {year}
  {2006})}\BibitemShut {NoStop}%
\bibitem [{\citenamefont {Blanchard}\ \emph {et~al.}(2016)\citenamefont
  {Blanchard}, \citenamefont {Hugonin},\ and\ \citenamefont
  {Sauvan}}]{Blanchard16}%
  \BibitemOpen
  \bibfield  {author} {\bibinfo {author} {\bibfnamefont {C.}~\bibnamefont
  {Blanchard}}, \bibinfo {author} {\bibfnamefont {J.-P.}\ \bibnamefont
  {Hugonin}},\ and\ \bibinfo {author} {\bibfnamefont {C.}~\bibnamefont
  {Sauvan}},\ }\bibfield  {title} {\bibinfo {title} {Fano resonances in
  photonic crystal slabs near optical bound states in the continuum},\ }\href
  {https://doi.org/10.1103/physrevb.94.155303} {\bibfield  {journal} {\bibinfo
  {journal} {Physical Review B}\ }\textbf {\bibinfo {volume} {94}},\ \bibinfo
  {pages} {155303} (\bibinfo {year} {2016})}\BibitemShut {NoStop}%
\bibitem [{\citenamefont {Bogdanov}\ \emph {et~al.}(2019)\citenamefont
  {Bogdanov}, \citenamefont {Koshelev}, \citenamefont {Kapitanova},
  \citenamefont {Rybin}, \citenamefont {Gladyshev}, \citenamefont {Sadrieva},
  \citenamefont {Samusev}, \citenamefont {Kivshar},\ and\ \citenamefont
  {Limonov}}]{bogdanov2019bound}%
  \BibitemOpen
  \bibfield  {author} {\bibinfo {author} {\bibfnamefont {A.~A.}\ \bibnamefont
  {Bogdanov}}, \bibinfo {author} {\bibfnamefont {K.~L.}\ \bibnamefont
  {Koshelev}}, \bibinfo {author} {\bibfnamefont {P.~V.}\ \bibnamefont
  {Kapitanova}}, \bibinfo {author} {\bibfnamefont {M.~V.}\ \bibnamefont
  {Rybin}}, \bibinfo {author} {\bibfnamefont {S.~A.}\ \bibnamefont
  {Gladyshev}}, \bibinfo {author} {\bibfnamefont {Z.~F.}\ \bibnamefont
  {Sadrieva}}, \bibinfo {author} {\bibfnamefont {K.~B.}\ \bibnamefont
  {Samusev}}, \bibinfo {author} {\bibfnamefont {Y.~S.}\ \bibnamefont
  {Kivshar}},\ and\ \bibinfo {author} {\bibfnamefont {M.~F.}\ \bibnamefont
  {Limonov}},\ }\bibfield  {title} {\bibinfo {title} {Bound states in the
  continuum and fano resonances in the strong mode coupling regime},\
  }\href@noop {} {\bibfield  {journal} {\bibinfo  {journal} {Advanced
  Photonics}\ }\textbf {\bibinfo {volume} {1}},\ \bibinfo {pages} {016001}
  (\bibinfo {year} {2019})}\BibitemShut {NoStop}%
\bibitem [{\citenamefont {Pankin}\ \emph {et~al.}(2020)\citenamefont {Pankin},
  \citenamefont {Maksimov}, \citenamefont {Chen},\ and\ \citenamefont
  {Timofeev}}]{pankin2020fano}%
  \BibitemOpen
  \bibfield  {author} {\bibinfo {author} {\bibfnamefont {P.~S.}\ \bibnamefont
  {Pankin}}, \bibinfo {author} {\bibfnamefont {D.~N.}\ \bibnamefont
  {Maksimov}}, \bibinfo {author} {\bibfnamefont {K.-P.}\ \bibnamefont {Chen}},\
  and\ \bibinfo {author} {\bibfnamefont {I.~V.}\ \bibnamefont {Timofeev}},\
  }\bibfield  {title} {\bibinfo {title} {Fano feature induced by a bound state
  in the continuum via resonant state expansion},\ }\href
  {https://doi.org/10.1038/s41598-020-70654-2} {\bibfield  {journal} {\bibinfo
  {journal} {Scientific Reports}\ }\textbf {\bibinfo {volume} {10}},\ \bibinfo
  {pages} {13691} (\bibinfo {year} {2020})}\BibitemShut {NoStop}%
\bibitem [{\citenamefont {Bulgakov}\ and\ \citenamefont
  {Maksimov}(2018)}]{Bulgakov18b}%
  \BibitemOpen
  \bibfield  {author} {\bibinfo {author} {\bibfnamefont {E.~N.}\ \bibnamefont
  {Bulgakov}}\ and\ \bibinfo {author} {\bibfnamefont {D.~N.}\ \bibnamefont
  {Maksimov}},\ }\bibfield  {title} {\bibinfo {title} {Optical response induced
  by bound states in the continuum in arrays of dielectric spheres},\ }\href
  {https://doi.org/10.1364/josab.35.002443} {\bibfield  {journal} {\bibinfo
  {journal} {Journal of the Optical Society of America B}\ }\textbf {\bibinfo
  {volume} {35}},\ \bibinfo {pages} {2443} (\bibinfo {year}
  {2018})}\BibitemShut {NoStop}%
\bibitem [{\citenamefont {Yoon}\ \emph {et~al.}(2015)\citenamefont {Yoon},
  \citenamefont {Song},\ and\ \citenamefont {Magnusson}}]{Yoon15}%
  \BibitemOpen
  \bibfield  {author} {\bibinfo {author} {\bibfnamefont {J.~W.}\ \bibnamefont
  {Yoon}}, \bibinfo {author} {\bibfnamefont {S.~H.}\ \bibnamefont {Song}},\
  and\ \bibinfo {author} {\bibfnamefont {R.}~\bibnamefont {Magnusson}},\
  }\bibfield  {title} {\bibinfo {title} {Critical field enhancement of
  asymptotic optical bound states in the continuum},\ }\href
  {https://doi.org/10.1038/srep18301} {\bibfield  {journal} {\bibinfo
  {journal} {Scientific Reports}\ }\textbf {\bibinfo {volume} {5}},\ \bibinfo
  {pages} {18301} (\bibinfo {year} {2015})}\BibitemShut {NoStop}%
\bibitem [{\citenamefont {Mocella}\ and\ \citenamefont
  {Romano}(2015)}]{Mocella15a}%
  \BibitemOpen
  \bibfield  {author} {\bibinfo {author} {\bibfnamefont {V.}~\bibnamefont
  {Mocella}}\ and\ \bibinfo {author} {\bibfnamefont {S.}~\bibnamefont
  {Romano}},\ }\bibfield  {title} {\bibinfo {title} {Giant field enhancement in
  photonic resonant lattices},\ }\href
  {https://doi.org/10.1103/physrevb.92.155117} {\bibfield  {journal} {\bibinfo
  {journal} {Physical Review B}\ }\textbf {\bibinfo {volume} {92}},\ \bibinfo
  {pages} {155117} (\bibinfo {year} {2015})}\BibitemShut {NoStop}%
\bibitem [{\citenamefont {Campione}\ \emph {et~al.}(2016)\citenamefont
  {Campione}, \citenamefont {Liu}, \citenamefont {Basilio}, \citenamefont
  {Warne}, \citenamefont {Langston}, \citenamefont {Luk}, \citenamefont
  {Wendt}, \citenamefont {Reno}, \citenamefont {Keeler}, \citenamefont
  {Brener},\ and\ \citenamefont {Sinclair}}]{Campione2016}%
  \BibitemOpen
  \bibfield  {author} {\bibinfo {author} {\bibfnamefont {S.}~\bibnamefont
  {Campione}}, \bibinfo {author} {\bibfnamefont {S.}~\bibnamefont {Liu}},
  \bibinfo {author} {\bibfnamefont {L.~I.}\ \bibnamefont {Basilio}}, \bibinfo
  {author} {\bibfnamefont {L.~K.}\ \bibnamefont {Warne}}, \bibinfo {author}
  {\bibfnamefont {W.~L.}\ \bibnamefont {Langston}}, \bibinfo {author}
  {\bibfnamefont {T.~S.}\ \bibnamefont {Luk}}, \bibinfo {author} {\bibfnamefont
  {J.~R.}\ \bibnamefont {Wendt}}, \bibinfo {author} {\bibfnamefont {J.~L.}\
  \bibnamefont {Reno}}, \bibinfo {author} {\bibfnamefont {G.~A.}\ \bibnamefont
  {Keeler}}, \bibinfo {author} {\bibfnamefont {I.}~\bibnamefont {Brener}},\
  and\ \bibinfo {author} {\bibfnamefont {M.~B.}\ \bibnamefont {Sinclair}},\
  }\bibfield  {title} {\bibinfo {title} {Broken symmetry dielectric resonators
  for high quality factor fano metasurfaces},\ }\href
  {https://doi.org/10.1021/acsphotonics.6b00556} {\bibfield  {journal}
  {\bibinfo  {journal} {ACS Photonics}\ }\textbf {\bibinfo {volume} {3}},\
  \bibinfo {pages} {2362} (\bibinfo {year} {2016})}\BibitemShut {NoStop}%
\bibitem [{\citenamefont {Zhou}\ \emph {et~al.}(2014)\citenamefont {Zhou},
  \citenamefont {Zhao}, \citenamefont {Shuai}, \citenamefont {Yang},
  \citenamefont {Chuwongin}, \citenamefont {Chadha}, \citenamefont {Seo},
  \citenamefont {Wang}, \citenamefont {Liu}, \citenamefont {Ma},\ and\
  \citenamefont {Fan}}]{Zhou2014}%
  \BibitemOpen
  \bibfield  {author} {\bibinfo {author} {\bibfnamefont {W.}~\bibnamefont
  {Zhou}}, \bibinfo {author} {\bibfnamefont {D.}~\bibnamefont {Zhao}}, \bibinfo
  {author} {\bibfnamefont {Y.-C.}\ \bibnamefont {Shuai}}, \bibinfo {author}
  {\bibfnamefont {H.}~\bibnamefont {Yang}}, \bibinfo {author} {\bibfnamefont
  {S.}~\bibnamefont {Chuwongin}}, \bibinfo {author} {\bibfnamefont
  {A.}~\bibnamefont {Chadha}}, \bibinfo {author} {\bibfnamefont {J.-H.}\
  \bibnamefont {Seo}}, \bibinfo {author} {\bibfnamefont {K.~X.}\ \bibnamefont
  {Wang}}, \bibinfo {author} {\bibfnamefont {V.}~\bibnamefont {Liu}}, \bibinfo
  {author} {\bibfnamefont {Z.}~\bibnamefont {Ma}},\ and\ \bibinfo {author}
  {\bibfnamefont {S.}~\bibnamefont {Fan}},\ }\bibfield  {title} {\bibinfo
  {title} {Progress in 2d photonic crystal fano resonance photonics},\
  }\href@noop {} {\bibfield  {journal} {\bibinfo  {journal} {Progress in
  Quantum Electronics}\ }\textbf {\bibinfo {volume} {38}},\ \bibinfo {pages}
  {1} (\bibinfo {year} {2014})}\BibitemShut {NoStop}%
\bibitem [{\citenamefont {Limonov}\ \emph {et~al.}(2017)\citenamefont
  {Limonov}, \citenamefont {Rybin}, \citenamefont {Poddubny},\ and\
  \citenamefont {Kivshar}}]{Limonov2017}%
  \BibitemOpen
  \bibfield  {author} {\bibinfo {author} {\bibfnamefont {M.~F.}\ \bibnamefont
  {Limonov}}, \bibinfo {author} {\bibfnamefont {M.~V.}\ \bibnamefont {Rybin}},
  \bibinfo {author} {\bibfnamefont {A.~N.}\ \bibnamefont {Poddubny}},\ and\
  \bibinfo {author} {\bibfnamefont {Y.~S.}\ \bibnamefont {Kivshar}},\
  }\bibfield  {title} {\bibinfo {title} {Fano resonances in photonics},\ }\href
  {https://doi.org/10.1038/nphoton.2017.142} {\bibfield  {journal} {\bibinfo
  {journal} {Nature Photonics}\ }\textbf {\bibinfo {volume} {11}},\ \bibinfo
  {pages} {543} (\bibinfo {year} {2017})}\BibitemShut {NoStop}%
\bibitem [{\citenamefont {Krasnok}\ \emph {et~al.}(2019)\citenamefont
  {Krasnok}, \citenamefont {Baranov}, \citenamefont {Li}, \citenamefont {Miri},
  \citenamefont {Monticone},\ and\ \citenamefont {Al{\'{u}}}}]{Krasnok19}%
  \BibitemOpen
  \bibfield  {author} {\bibinfo {author} {\bibfnamefont {A.}~\bibnamefont
  {Krasnok}}, \bibinfo {author} {\bibfnamefont {D.}~\bibnamefont {Baranov}},
  \bibinfo {author} {\bibfnamefont {H.}~\bibnamefont {Li}}, \bibinfo {author}
  {\bibfnamefont {M.-A.}\ \bibnamefont {Miri}}, \bibinfo {author}
  {\bibfnamefont {F.}~\bibnamefont {Monticone}},\ and\ \bibinfo {author}
  {\bibfnamefont {A.}~\bibnamefont {Al{\'{u}}}},\ }\bibfield  {title} {\bibinfo
  {title} {Anomalies in light scattering},\ }\href
  {https://doi.org/10.1364/aop.11.000892} {\bibfield  {journal} {\bibinfo
  {journal} {Advances in Optics and Photonics}\ }\textbf {\bibinfo {volume}
  {11}},\ \bibinfo {pages} {892} (\bibinfo {year} {2019})}\BibitemShut
  {NoStop}%
\bibitem [{\citenamefont {Fan}\ \emph {et~al.}(2003)\citenamefont {Fan},
  \citenamefont {Suh},\ and\ \citenamefont {Joannopoulos}}]{Fan03}%
  \BibitemOpen
  \bibfield  {author} {\bibinfo {author} {\bibfnamefont {S.}~\bibnamefont
  {Fan}}, \bibinfo {author} {\bibfnamefont {W.}~\bibnamefont {Suh}},\ and\
  \bibinfo {author} {\bibfnamefont {J.~D.}\ \bibnamefont {Joannopoulos}},\
  }\bibfield  {title} {\bibinfo {title} {Temporal coupled-mode theory for the
  fano resonance in optical resonators},\ }\href
  {https://doi.org/10.1364/josaa.20.000569} {\bibfield  {journal} {\bibinfo
  {journal} {Journal of the Optical Society of America A}\ }\textbf {\bibinfo
  {volume} {20}},\ \bibinfo {pages} {569} (\bibinfo {year} {2003})}\BibitemShut
  {NoStop}%
\bibitem [{\citenamefont {Alpeggiani}\ \emph {et~al.}(2017)\citenamefont
  {Alpeggiani}, \citenamefont {Parappurath}, \citenamefont {Verhagen},\ and\
  \citenamefont {Kuipers}}]{Alpeggiani2017}%
  \BibitemOpen
  \bibfield  {author} {\bibinfo {author} {\bibfnamefont {F.}~\bibnamefont
  {Alpeggiani}}, \bibinfo {author} {\bibfnamefont {N.}~\bibnamefont
  {Parappurath}}, \bibinfo {author} {\bibfnamefont {E.}~\bibnamefont
  {Verhagen}},\ and\ \bibinfo {author} {\bibfnamefont {L.}~\bibnamefont
  {Kuipers}},\ }\bibfield  {title} {\bibinfo {title} {Quasinormal-mode
  expansion of the scattering matrix},\ }\href
  {https://doi.org/10.1103/physrevx.7.021035} {\bibfield  {journal} {\bibinfo
  {journal} {Physical Review X}\ }\textbf {\bibinfo {volume} {7}},\ \bibinfo
  {pages} {021035} (\bibinfo {year} {2017})}\BibitemShut {NoStop}%
\bibitem [{\citenamefont {Ming}\ \emph {et~al.}(2017)\citenamefont {Ming},
  \citenamefont {Liu}, \citenamefont {Sun},\ and\ \citenamefont
  {Padilla}}]{Ming2017}%
  \BibitemOpen
  \bibfield  {author} {\bibinfo {author} {\bibfnamefont {X.}~\bibnamefont
  {Ming}}, \bibinfo {author} {\bibfnamefont {X.}~\bibnamefont {Liu}}, \bibinfo
  {author} {\bibfnamefont {L.}~\bibnamefont {Sun}},\ and\ \bibinfo {author}
  {\bibfnamefont {W.~J.}\ \bibnamefont {Padilla}},\ }\bibfield  {title}
  {\bibinfo {title} {Degenerate critical coupling in all-dielectric metasurface
  absorbers},\ }\href {https://doi.org/10.1364/oe.25.024658} {\bibfield
  {journal} {\bibinfo  {journal} {Optics Express}\ }\textbf {\bibinfo {volume}
  {25}},\ \bibinfo {pages} {24658} (\bibinfo {year} {2017})}\BibitemShut
  {NoStop}%
\bibitem [{\citenamefont {Zhou}\ \emph {et~al.}(2016)\citenamefont {Zhou},
  \citenamefont {Zhen}, \citenamefont {Hsu}, \citenamefont {Miller},
  \citenamefont {Johnson}, \citenamefont {Joannopoulos},\ and\ \citenamefont
  {Solja{\v{c}}i{\'{c}}}}]{Zhou2016}%
  \BibitemOpen
  \bibfield  {author} {\bibinfo {author} {\bibfnamefont {H.}~\bibnamefont
  {Zhou}}, \bibinfo {author} {\bibfnamefont {B.}~\bibnamefont {Zhen}}, \bibinfo
  {author} {\bibfnamefont {C.~W.}\ \bibnamefont {Hsu}}, \bibinfo {author}
  {\bibfnamefont {O.~D.}\ \bibnamefont {Miller}}, \bibinfo {author}
  {\bibfnamefont {S.~G.}\ \bibnamefont {Johnson}}, \bibinfo {author}
  {\bibfnamefont {J.~D.}\ \bibnamefont {Joannopoulos}},\ and\ \bibinfo {author}
  {\bibfnamefont {M.}~\bibnamefont {Solja{\v{c}}i{\'{c}}}},\ }\bibfield
  {title} {\bibinfo {title} {Perfect single-sided radiation and absorption
  without mirrors},\ }\href {https://doi.org/10.1364/optica.3.001079}
  {\bibfield  {journal} {\bibinfo  {journal} {Optica}\ }\textbf {\bibinfo
  {volume} {3}},\ \bibinfo {pages} {1079} (\bibinfo {year} {2016})}\BibitemShut
  {NoStop}%
\bibitem [{\citenamefont {Maksimov}\ \emph
  {et~al.}(2020{\natexlab{b}})\citenamefont {Maksimov}, \citenamefont
  {Bogdanov},\ and\ \citenamefont {Bulgakov}}]{maksimov2020optical}%
  \BibitemOpen
  \bibfield  {author} {\bibinfo {author} {\bibfnamefont {D.~N.}\ \bibnamefont
  {Maksimov}}, \bibinfo {author} {\bibfnamefont {A.~A.}\ \bibnamefont
  {Bogdanov}},\ and\ \bibinfo {author} {\bibfnamefont {E.~N.}\ \bibnamefont
  {Bulgakov}},\ }\bibfield  {title} {\bibinfo {title} {Optical bistability with
  bound states in the continuum in dielectric gratings},\ }\href@noop {}
  {\bibfield  {journal} {\bibinfo  {journal} {Physical Review A}\ }\textbf
  {\bibinfo {volume} {102}},\ \bibinfo {pages} {033511} (\bibinfo {year}
  {2020}{\natexlab{b}})}\BibitemShut {NoStop}%
\bibitem [{\citenamefont {Bikbaev}\ \emph {et~al.}(2021)\citenamefont
  {Bikbaev}, \citenamefont {Maksimov}, \citenamefont {Pankin}, \citenamefont
  {Chen},\ and\ \citenamefont {Timofeev}}]{Bikbaev21}%
  \BibitemOpen
  \bibfield  {author} {\bibinfo {author} {\bibfnamefont {R.~G.}\ \bibnamefont
  {Bikbaev}}, \bibinfo {author} {\bibfnamefont {D.~N.}\ \bibnamefont
  {Maksimov}}, \bibinfo {author} {\bibfnamefont {P.~S.}\ \bibnamefont
  {Pankin}}, \bibinfo {author} {\bibfnamefont {K.-P.}\ \bibnamefont {Chen}},\
  and\ \bibinfo {author} {\bibfnamefont {I.~V.}\ \bibnamefont {Timofeev}},\
  }\bibfield  {title} {\bibinfo {title} {Critical coupling vortex with
  grating-induced high q-factor optical tamm states},\ }\href
  {https://doi.org/10.1364/oe.416132} {\bibfield  {journal} {\bibinfo
  {journal} {Optics Express}\ }\textbf {\bibinfo {volume} {29}},\ \bibinfo
  {pages} {4672} (\bibinfo {year} {2021})}\BibitemShut {NoStop}%
\bibitem [{\citenamefont {Zhang}\ \emph {et~al.}(2023)\citenamefont {Zhang},
  \citenamefont {You}, \citenamefont {Feng}, \citenamefont {Na}, \citenamefont
  {Lou}, \citenamefont {Zhang},\ and\ \citenamefont {Cui}}]{Zhang2023}%
  \BibitemOpen
  \bibfield  {author} {\bibinfo {author} {\bibfnamefont {J.}~\bibnamefont
  {Zhang}}, \bibinfo {author} {\bibfnamefont {J.~W.}\ \bibnamefont {You}},
  \bibinfo {author} {\bibfnamefont {F.}~\bibnamefont {Feng}}, \bibinfo {author}
  {\bibfnamefont {W.}~\bibnamefont {Na}}, \bibinfo {author} {\bibfnamefont
  {Z.~C.}\ \bibnamefont {Lou}}, \bibinfo {author} {\bibfnamefont {Q.-J.}\
  \bibnamefont {Zhang}},\ and\ \bibinfo {author} {\bibfnamefont {T.~J.}\
  \bibnamefont {Cui}},\ }\bibfield  {title} {\bibinfo {title} {Physics-driven
  machine-learning approach incorporating temporal coupled mode theory for
  intelligent design of metasurfaces},\ }\href
  {https://doi.org/10.1109/tmtt.2023.3238076} {\bibfield  {journal} {\bibinfo
  {journal} {IEEE Transactions on Microwave Theory and Techniques}\ }\textbf
  {\bibinfo {volume} {71}},\ \bibinfo {pages} {2875} (\bibinfo {year}
  {2023})}\BibitemShut {NoStop}%
\bibitem [{\citenamefont {Wu}\ \emph {et~al.}(2022)\citenamefont {Wu},
  \citenamefont {Yuan},\ and\ \citenamefont {Lu}}]{Wu2022}%
  \BibitemOpen
  \bibfield  {author} {\bibinfo {author} {\bibfnamefont {H.}~\bibnamefont
  {Wu}}, \bibinfo {author} {\bibfnamefont {L.}~\bibnamefont {Yuan}},\ and\
  \bibinfo {author} {\bibfnamefont {Y.~Y.}\ \bibnamefont {Lu}},\ }\bibfield
  {title} {\bibinfo {title} {Approximating transmission and reflection spectra
  near isolated nondegenerate resonances},\ }\href
  {https://doi.org/10.1103/physreva.105.063510} {\bibfield  {journal} {\bibinfo
   {journal} {Physical Review A}\ }\textbf {\bibinfo {volume} {105}},\ \bibinfo
  {pages} {063510} (\bibinfo {year} {2022})}\BibitemShut {NoStop}%
\bibitem [{\citenamefont {Huang}\ \emph {et~al.}(2024)\citenamefont {Huang},
  \citenamefont {Wang}, \citenamefont {Jia}, \citenamefont {Zhang},\ and\
  \citenamefont {Zhou}}]{Huang2024}%
  \BibitemOpen
  \bibfield  {author} {\bibinfo {author} {\bibfnamefont {Z.}~\bibnamefont
  {Huang}}, \bibinfo {author} {\bibfnamefont {J.}~\bibnamefont {Wang}},
  \bibinfo {author} {\bibfnamefont {W.}~\bibnamefont {Jia}}, \bibinfo {author}
  {\bibfnamefont {S.}~\bibnamefont {Zhang}},\ and\ \bibinfo {author}
  {\bibfnamefont {C.}~\bibnamefont {Zhou}},\ }\bibfield  {title} {\bibinfo
  {title} {All-dielectric metasurfaces enabled by quasi-bic for high-q
  near-perfect light absorption},\ }\href {https://doi.org/10.1364/ol.541553}
  {\bibfield  {journal} {\bibinfo  {journal} {Optics Letters}\ }\textbf
  {\bibinfo {volume} {50}},\ \bibinfo {pages} {105} (\bibinfo {year}
  {2024})}\BibitemShut {NoStop}%
\bibitem [{\citenamefont {Popov}\ \emph {et~al.}(1986)\citenamefont {Popov},
  \citenamefont {Mashev},\ and\ \citenamefont {Maystre}}]{Popov1986}%
  \BibitemOpen
  \bibfield  {author} {\bibinfo {author} {\bibfnamefont {E.}~\bibnamefont
  {Popov}}, \bibinfo {author} {\bibfnamefont {L.}~\bibnamefont {Mashev}},\ and\
  \bibinfo {author} {\bibfnamefont {D.}~\bibnamefont {Maystre}},\ }\bibfield
  {title} {\bibinfo {title} {Theoretical study of the anomalies of coated
  dielectric gratings},\ }\href {https://doi.org/10.1080/713821994} {\bibfield
  {journal} {\bibinfo  {journal} {Optica Acta: International Journal of
  Optics}\ }\textbf {\bibinfo {volume} {33}},\ \bibinfo {pages} {607} (\bibinfo
  {year} {1986})}\BibitemShut {NoStop}%
\bibitem [{\citenamefont {Shipman}\ and\ \citenamefont
  {Tu}(2012)}]{Shipman2012}%
  \BibitemOpen
  \bibfield  {author} {\bibinfo {author} {\bibfnamefont {S.~P.}\ \bibnamefont
  {Shipman}}\ and\ \bibinfo {author} {\bibfnamefont {H.}~\bibnamefont {Tu}},\
  }\bibfield  {title} {\bibinfo {title} {Total resonant transmission and
  reflection by periodic structures},\ }\href
  {https://doi.org/10.1137/110834196} {\bibfield  {journal} {\bibinfo
  {journal} {SIAM Journal on Applied Mathematics}\ }\textbf {\bibinfo {volume}
  {72}},\ \bibinfo {pages} {216} (\bibinfo {year} {2012})}\BibitemShut
  {NoStop}%
\bibitem [{\citenamefont {Wang}\ \emph {et~al.}(2013)\citenamefont {Wang},
  \citenamefont {Yu}, \citenamefont {Sandhu},\ and\ \citenamefont
  {Fan}}]{Wang2013}%
  \BibitemOpen
  \bibfield  {author} {\bibinfo {author} {\bibfnamefont {K.~X.}\ \bibnamefont
  {Wang}}, \bibinfo {author} {\bibfnamefont {Z.}~\bibnamefont {Yu}}, \bibinfo
  {author} {\bibfnamefont {S.}~\bibnamefont {Sandhu}},\ and\ \bibinfo {author}
  {\bibfnamefont {S.}~\bibnamefont {Fan}},\ }\bibfield  {title} {\bibinfo
  {title} {Fundamental bounds on decay rates in asymmetric single-mode optical
  resonators},\ }\href {https://doi.org/10.1364/ol.38.000100} {\bibfield
  {journal} {\bibinfo  {journal} {Optics Letters}\ }\textbf {\bibinfo {volume}
  {38}},\ \bibinfo {pages} {100} (\bibinfo {year} {2013})}\BibitemShut
  {NoStop}%
\bibitem [{\citenamefont {Bykov}\ and\ \citenamefont
  {Doskolovich}(2015)}]{Bykov15}%
  \BibitemOpen
  \bibfield  {author} {\bibinfo {author} {\bibfnamefont {D.~A.}\ \bibnamefont
  {Bykov}}\ and\ \bibinfo {author} {\bibfnamefont {L.~L.}\ \bibnamefont
  {Doskolovich}},\ }\bibfield  {title} {\bibinfo {title} {$\omega$- $k_x$
  {Fano} line shape in photonic crystal slabs},\ }\href
  {https://doi.org/10.1103/physreva.92.013845} {\bibfield  {journal} {\bibinfo
  {journal} {Physical Review A}\ }\textbf {\bibinfo {volume} {92}},\ \bibinfo
  {pages} {013845} (\bibinfo {year} {2015})}\BibitemShut {NoStop}%
\bibitem [{\citenamefont {Yuan}\ \emph {et~al.}(2022)\citenamefont {Yuan},
  \citenamefont {Zhang},\ and\ \citenamefont {Lu}}]{Yuan2022}%
  \BibitemOpen
  \bibfield  {author} {\bibinfo {author} {\bibfnamefont {L.}~\bibnamefont
  {Yuan}}, \bibinfo {author} {\bibfnamefont {M.}~\bibnamefont {Zhang}},\ and\
  \bibinfo {author} {\bibfnamefont {Y.~Y.}\ \bibnamefont {Lu}},\ }\bibfield
  {title} {\bibinfo {title} {Real transmission and reflection zeros of periodic
  structures with a bound state in the continuum},\ }\href
  {https://doi.org/10.1103/physreva.106.013505} {\bibfield  {journal} {\bibinfo
   {journal} {Physical Review A}\ }\textbf {\bibinfo {volume} {106}},\ \bibinfo
  {pages} {013505} (\bibinfo {year} {2022})}\BibitemShut {NoStop}%
\bibitem [{\citenamefont {Ma}\ \emph {et~al.}(2021)\citenamefont {Ma},
  \citenamefont {Liu}, \citenamefont {Kudyshev}, \citenamefont {Boltasseva},
  \citenamefont {Cai},\ and\ \citenamefont {Liu}}]{ma2021deep}%
  \BibitemOpen
  \bibfield  {author} {\bibinfo {author} {\bibfnamefont {W.}~\bibnamefont
  {Ma}}, \bibinfo {author} {\bibfnamefont {Z.}~\bibnamefont {Liu}}, \bibinfo
  {author} {\bibfnamefont {Z.~A.}\ \bibnamefont {Kudyshev}}, \bibinfo {author}
  {\bibfnamefont {A.}~\bibnamefont {Boltasseva}}, \bibinfo {author}
  {\bibfnamefont {W.}~\bibnamefont {Cai}},\ and\ \bibinfo {author}
  {\bibfnamefont {Y.}~\bibnamefont {Liu}},\ }\bibfield  {title} {\bibinfo
  {title} {Deep learning for the design of photonic structures},\ }\href@noop
  {} {\bibfield  {journal} {\bibinfo  {journal} {Nature Photonics}\ }\textbf
  {\bibinfo {volume} {15}},\ \bibinfo {pages} {77} (\bibinfo {year}
  {2021})}\BibitemShut {NoStop}%
\bibitem [{\citenamefont {Jiang}\ \emph {et~al.}(2021)\citenamefont {Jiang},
  \citenamefont {Chen},\ and\ \citenamefont {Fan}}]{jiang2021deep}%
  \BibitemOpen
  \bibfield  {author} {\bibinfo {author} {\bibfnamefont {J.}~\bibnamefont
  {Jiang}}, \bibinfo {author} {\bibfnamefont {M.}~\bibnamefont {Chen}},\ and\
  \bibinfo {author} {\bibfnamefont {J.~A.}\ \bibnamefont {Fan}},\ }\bibfield
  {title} {\bibinfo {title} {Deep neural networks for the evaluation and design
  of photonic devices},\ }\href@noop {} {\bibfield  {journal} {\bibinfo
  {journal} {Nature Reviews Materials}\ }\textbf {\bibinfo {volume} {6}},\
  \bibinfo {pages} {679} (\bibinfo {year} {2021})}\BibitemShut {NoStop}%
\bibitem [{\citenamefont {So}\ \emph {et~al.}(2020)\citenamefont {So},
  \citenamefont {Badloe}, \citenamefont {Noh}, \citenamefont {Bravo-Abad},\
  and\ \citenamefont {Rho}}]{so2020deep}%
  \BibitemOpen
  \bibfield  {author} {\bibinfo {author} {\bibfnamefont {S.}~\bibnamefont
  {So}}, \bibinfo {author} {\bibfnamefont {T.}~\bibnamefont {Badloe}}, \bibinfo
  {author} {\bibfnamefont {J.}~\bibnamefont {Noh}}, \bibinfo {author}
  {\bibfnamefont {J.}~\bibnamefont {Bravo-Abad}},\ and\ \bibinfo {author}
  {\bibfnamefont {J.}~\bibnamefont {Rho}},\ }\bibfield  {title} {\bibinfo
  {title} {Deep learning enabled inverse design in nanophotonics},\ }\href@noop
  {} {\bibfield  {journal} {\bibinfo  {journal} {Nanophotonics}\ }\textbf
  {\bibinfo {volume} {9}},\ \bibinfo {pages} {1041} (\bibinfo {year}
  {2020})}\BibitemShut {NoStop}%
\bibitem [{\citenamefont {Pilozzi}\ \emph {et~al.}(2018)\citenamefont
  {Pilozzi}, \citenamefont {Farrelly}, \citenamefont {Marcucci},\ and\
  \citenamefont {Conti}}]{pilozzi2018machine}%
  \BibitemOpen
  \bibfield  {author} {\bibinfo {author} {\bibfnamefont {L.}~\bibnamefont
  {Pilozzi}}, \bibinfo {author} {\bibfnamefont {F.~A.}\ \bibnamefont
  {Farrelly}}, \bibinfo {author} {\bibfnamefont {G.}~\bibnamefont {Marcucci}},\
  and\ \bibinfo {author} {\bibfnamefont {C.}~\bibnamefont {Conti}},\ }\bibfield
   {title} {\bibinfo {title} {Machine learning inverse problem for topological
  photonics},\ }\href@noop {} {\bibfield  {journal} {\bibinfo  {journal}
  {Communications Physics}\ }\textbf {\bibinfo {volume} {1}},\ \bibinfo {pages}
  {57} (\bibinfo {year} {2018})}\BibitemShut {NoStop}%
\bibitem [{\citenamefont {Kudyshev}\ \emph {et~al.}(2020)\citenamefont
  {Kudyshev}, \citenamefont {Shalaev},\ and\ \citenamefont
  {Boltasseva}}]{kudyshev2020machine}%
  \BibitemOpen
  \bibfield  {author} {\bibinfo {author} {\bibfnamefont {Z.~A.}\ \bibnamefont
  {Kudyshev}}, \bibinfo {author} {\bibfnamefont {V.~M.}\ \bibnamefont
  {Shalaev}},\ and\ \bibinfo {author} {\bibfnamefont {A.}~\bibnamefont
  {Boltasseva}},\ }\bibfield  {title} {\bibinfo {title} {Machine learning for
  integrated quantum photonics},\ }\href@noop {} {\bibfield  {journal}
  {\bibinfo  {journal} {Acs Photonics}\ }\textbf {\bibinfo {volume} {8}},\
  \bibinfo {pages} {34} (\bibinfo {year} {2020})}\BibitemShut {NoStop}%
\bibitem [{\citenamefont {Zhao}\ \emph {et~al.}(2023)\citenamefont {Zhao},
  \citenamefont {Qing}, \citenamefont {Kong}, \citenamefont {Xu}, \citenamefont
  {Fan}, \citenamefont {Yun}, \citenamefont {Zhang},\ and\ \citenamefont
  {Wu}}]{Zhao2023}%
  \BibitemOpen
  \bibfield  {author} {\bibinfo {author} {\bibfnamefont {Z.}~\bibnamefont
  {Zhao}}, \bibinfo {author} {\bibfnamefont {Y.}~\bibnamefont {Qing}}, \bibinfo
  {author} {\bibfnamefont {L.}~\bibnamefont {Kong}}, \bibinfo {author}
  {\bibfnamefont {H.}~\bibnamefont {Xu}}, \bibinfo {author} {\bibfnamefont
  {X.}~\bibnamefont {Fan}}, \bibinfo {author} {\bibfnamefont {J.}~\bibnamefont
  {Yun}}, \bibinfo {author} {\bibfnamefont {L.}~\bibnamefont {Zhang}},\ and\
  \bibinfo {author} {\bibfnamefont {H.}~\bibnamefont {Wu}},\ }\bibfield
  {title} {\bibinfo {title} {Advancements in microwave absorption motivated by
  interdisciplinary research},\ }\bibfield  {journal} {\bibinfo  {journal}
  {Advanced Materials}\ }\textbf {\bibinfo {volume} {36}},\ \href
  {https://doi.org/10.1002/adma.202304182} {10.1002/adma.202304182} (\bibinfo
  {year} {2023})\BibitemShut {NoStop}%
\bibitem [{\citenamefont {Deng}\ \emph {et~al.}(2025)\citenamefont {Deng},
  \citenamefont {Fan}, \citenamefont {Jin}, \citenamefont {Malof},\ and\
  \citenamefont {Padilla}}]{Deng2025}%
  \BibitemOpen
  \bibfield  {author} {\bibinfo {author} {\bibfnamefont {Y.}~\bibnamefont
  {Deng}}, \bibinfo {author} {\bibfnamefont {K.}~\bibnamefont {Fan}}, \bibinfo
  {author} {\bibfnamefont {B.}~\bibnamefont {Jin}}, \bibinfo {author}
  {\bibfnamefont {J.}~\bibnamefont {Malof}},\ and\ \bibinfo {author}
  {\bibfnamefont {W.~J.}\ \bibnamefont {Padilla}},\ }\bibfield  {title}
  {\bibinfo {title} {Physics-informed learning in artificial electromagnetic
  materials},\ }\bibfield  {journal} {\bibinfo  {journal} {Applied Physics
  Reviews}\ }\textbf {\bibinfo {volume} {12}},\ \href
  {https://doi.org/10.1063/5.0232675} {10.1063/5.0232675} (\bibinfo {year}
  {2025})\BibitemShut {NoStop}%
\bibitem [{\citenamefont {Lin}\ \emph {et~al.}(2021)\citenamefont {Lin},
  \citenamefont {Alnakhli},\ and\ \citenamefont {Li}}]{lin2021engineering}%
  \BibitemOpen
  \bibfield  {author} {\bibinfo {author} {\bibfnamefont {R.}~\bibnamefont
  {Lin}}, \bibinfo {author} {\bibfnamefont {Z.}~\bibnamefont {Alnakhli}},\ and\
  \bibinfo {author} {\bibfnamefont {X.}~\bibnamefont {Li}},\ }\bibfield
  {title} {\bibinfo {title} {Engineering of multiple bound states in the
  continuum by latent representation of freeform structures},\ }\href@noop {}
  {\bibfield  {journal} {\bibinfo  {journal} {Photonics Research}\ }\textbf
  {\bibinfo {volume} {9}},\ \bibinfo {pages} {B96} (\bibinfo {year}
  {2021})}\BibitemShut {NoStop}%
\bibitem [{\citenamefont {Ma}\ \emph {et~al.}(2022)\citenamefont {Ma},
  \citenamefont {Ma}, \citenamefont {Cunha}, \citenamefont {Liu}, \citenamefont
  {Kudtarkar}, \citenamefont {Xu}, \citenamefont {Wang}, \citenamefont {Chen},
  \citenamefont {Wong}, \citenamefont {Liu} \emph
  {et~al.}}]{ma2022strategical}%
  \BibitemOpen
  \bibfield  {author} {\bibinfo {author} {\bibfnamefont {X.}~\bibnamefont
  {Ma}}, \bibinfo {author} {\bibfnamefont {Y.}~\bibnamefont {Ma}}, \bibinfo
  {author} {\bibfnamefont {P.}~\bibnamefont {Cunha}}, \bibinfo {author}
  {\bibfnamefont {Q.}~\bibnamefont {Liu}}, \bibinfo {author} {\bibfnamefont
  {K.}~\bibnamefont {Kudtarkar}}, \bibinfo {author} {\bibfnamefont
  {D.}~\bibnamefont {Xu}}, \bibinfo {author} {\bibfnamefont {J.}~\bibnamefont
  {Wang}}, \bibinfo {author} {\bibfnamefont {Y.}~\bibnamefont {Chen}}, \bibinfo
  {author} {\bibfnamefont {Z.~J.}\ \bibnamefont {Wong}}, \bibinfo {author}
  {\bibfnamefont {M.}~\bibnamefont {Liu}}, \emph {et~al.},\ }\bibfield  {title}
  {\bibinfo {title} {Strategical deep learning for photonic bound states in the
  continuum},\ }\href@noop {} {\bibfield  {journal} {\bibinfo  {journal} {Laser
  \& Photonics Reviews}\ }\textbf {\bibinfo {volume} {16}},\ \bibinfo {pages}
  {2100658} (\bibinfo {year} {2022})}\BibitemShut {NoStop}%
\bibitem [{\citenamefont {Wang}\ \emph
  {et~al.}(2023{\natexlab{a}})\citenamefont {Wang}, \citenamefont {Chen},
  \citenamefont {Zhang}, \citenamefont {Zhang}, \citenamefont {Zhou},
  \citenamefont {Zuo}, \citenamefont {Chen},\ and\ \citenamefont
  {Peng}}]{wang2023automatic}%
  \BibitemOpen
  \bibfield  {author} {\bibinfo {author} {\bibfnamefont {F.}~\bibnamefont
  {Wang}}, \bibinfo {author} {\bibfnamefont {Y.}~\bibnamefont {Chen}}, \bibinfo
  {author} {\bibfnamefont {Z.}~\bibnamefont {Zhang}}, \bibinfo {author}
  {\bibfnamefont {X.}~\bibnamefont {Zhang}}, \bibinfo {author} {\bibfnamefont
  {X.}~\bibnamefont {Zhou}}, \bibinfo {author} {\bibfnamefont {Y.}~\bibnamefont
  {Zuo}}, \bibinfo {author} {\bibfnamefont {Z.}~\bibnamefont {Chen}},\ and\
  \bibinfo {author} {\bibfnamefont {C.}~\bibnamefont {Peng}},\ }\bibfield
  {title} {\bibinfo {title} {Automatic optimization of miniaturized bound
  states in the continuum cavity},\ }\href@noop {} {\bibfield  {journal}
  {\bibinfo  {journal} {Optics Express}\ }\textbf {\bibinfo {volume} {31}},\
  \bibinfo {pages} {12384} (\bibinfo {year} {2023}{\natexlab{a}})}\BibitemShut
  {NoStop}%
\bibitem [{\citenamefont {Wang}\ \emph
  {et~al.}(2023{\natexlab{b}})\citenamefont {Wang}, \citenamefont {Sun},
  \citenamefont {Li}, \citenamefont {Wang}, \citenamefont {Li}, \citenamefont
  {Zheng},\ and\ \citenamefont {Wen}}]{Wang2023}%
  \BibitemOpen
  \bibfield  {author} {\bibinfo {author} {\bibfnamefont {Z.}~\bibnamefont
  {Wang}}, \bibinfo {author} {\bibfnamefont {J.}~\bibnamefont {Sun}}, \bibinfo
  {author} {\bibfnamefont {J.}~\bibnamefont {Li}}, \bibinfo {author}
  {\bibfnamefont {L.}~\bibnamefont {Wang}}, \bibinfo {author} {\bibfnamefont
  {Z.}~\bibnamefont {Li}}, \bibinfo {author} {\bibfnamefont {X.}~\bibnamefont
  {Zheng}},\ and\ \bibinfo {author} {\bibfnamefont {L.}~\bibnamefont {Wen}},\
  }\bibfield  {title} {\bibinfo {title} {Customizing 2.5d out-of-plane
  architectures for robust plasmonic bound-states-in-the-continuum
  metasurfaces},\ }\href {https://doi.org/10.1002/advs.202206236} {\bibfield
  {journal} {\bibinfo  {journal} {Advanced Science}\ }\textbf {\bibinfo
  {volume} {10}},\ \bibinfo {pages} {2206236} (\bibinfo {year}
  {2023}{\natexlab{b}})}\BibitemShut {NoStop}%
\bibitem [{\citenamefont {Zhang}\ \emph {et~al.}(2024)\citenamefont {Zhang},
  \citenamefont {Chen}, \citenamefont {Lin}, \citenamefont {Yu}, \citenamefont
  {Ma}, \citenamefont {Lu}, \citenamefont {You},\ and\ \citenamefont
  {Cui}}]{Zhang2024}%
  \BibitemOpen
  \bibfield  {author} {\bibinfo {author} {\bibfnamefont {J.~N.}\ \bibnamefont
  {Zhang}}, \bibinfo {author} {\bibfnamefont {L.}~\bibnamefont {Chen}},
  \bibinfo {author} {\bibfnamefont {X.~M.}\ \bibnamefont {Lin}}, \bibinfo
  {author} {\bibfnamefont {X.~Y.}\ \bibnamefont {Yu}}, \bibinfo {author}
  {\bibfnamefont {Q.}~\bibnamefont {Ma}}, \bibinfo {author} {\bibfnamefont
  {W.-B.}\ \bibnamefont {Lu}}, \bibinfo {author} {\bibfnamefont {J.~W.}\
  \bibnamefont {You}},\ and\ \bibinfo {author} {\bibfnamefont {T.~J.}\
  \bibnamefont {Cui}},\ }\bibfield  {title} {\bibinfo {title} {Feature-assisted
  neuro-cmt approach to fast design optimization of metasurfaces},\ }\href
  {https://doi.org/10.1109/lmwt.2024.3381112} {\bibfield  {journal} {\bibinfo
  {journal} {IEEE Microwave and Wireless Technology Letters}\ }\textbf
  {\bibinfo {volume} {34}},\ \bibinfo {pages} {467} (\bibinfo {year}
  {2024})}\BibitemShut {NoStop}%
\bibitem [{\citenamefont {Su}\ \emph {et~al.}(2024)\citenamefont {Su},
  \citenamefont {You}, \citenamefont {Chen}, \citenamefont {Yu}, \citenamefont
  {Yin}, \citenamefont {Yuan}, \citenamefont {Huang}, \citenamefont {Ma},
  \citenamefont {Zhang},\ and\ \citenamefont {Cui}}]{Su2024}%
  \BibitemOpen
  \bibfield  {author} {\bibinfo {author} {\bibfnamefont {J.~L.}\ \bibnamefont
  {Su}}, \bibinfo {author} {\bibfnamefont {J.~W.}\ \bibnamefont {You}},
  \bibinfo {author} {\bibfnamefont {L.}~\bibnamefont {Chen}}, \bibinfo {author}
  {\bibfnamefont {X.~Y.}\ \bibnamefont {Yu}}, \bibinfo {author} {\bibfnamefont
  {Q.~C.}\ \bibnamefont {Yin}}, \bibinfo {author} {\bibfnamefont {G.~H.}\
  \bibnamefont {Yuan}}, \bibinfo {author} {\bibfnamefont {S.~Q.}\ \bibnamefont
  {Huang}}, \bibinfo {author} {\bibfnamefont {Q.}~\bibnamefont {Ma}}, \bibinfo
  {author} {\bibfnamefont {J.~N.}\ \bibnamefont {Zhang}},\ and\ \bibinfo
  {author} {\bibfnamefont {T.~J.}\ \bibnamefont {Cui}},\ }\bibfield  {title}
  {\bibinfo {title} {Metaphynet: intelligent design of large-scale metasurfaces
  based on physics-driven neural network},\ }\href
  {https://doi.org/10.1088/2515-7647/ad4cc8} {\bibfield  {journal} {\bibinfo
  {journal} {Journal of Physics: Photonics}\ }\textbf {\bibinfo {volume} {6}},\
  \bibinfo {pages} {035010} (\bibinfo {year} {2024})}\BibitemShut {NoStop}%
\bibitem [{\citenamefont {Molokeev}\ \emph {et~al.}(2023)\citenamefont
  {Molokeev}, \citenamefont {Kostyukov}, \citenamefont {Ershov}, \citenamefont
  {Maksimov}, \citenamefont {Gerasimov},\ and\ \citenamefont
  {Polyutov}}]{Molokeev2023}%
  \BibitemOpen
  \bibfield  {author} {\bibinfo {author} {\bibfnamefont {M.~S.}\ \bibnamefont
  {Molokeev}}, \bibinfo {author} {\bibfnamefont {A.~S.}\ \bibnamefont
  {Kostyukov}}, \bibinfo {author} {\bibfnamefont {A.~E.}\ \bibnamefont
  {Ershov}}, \bibinfo {author} {\bibfnamefont {D.~N.}\ \bibnamefont
  {Maksimov}}, \bibinfo {author} {\bibfnamefont {V.~S.}\ \bibnamefont
  {Gerasimov}},\ and\ \bibinfo {author} {\bibfnamefont {S.~P.}\ \bibnamefont
  {Polyutov}},\ }\bibfield  {title} {\bibinfo {title} {Infrared bound states in
  the continuum: random forest method},\ }\href
  {https://doi.org/10.1364/ol.494629} {\bibfield  {journal} {\bibinfo
  {journal} {Optics Letters}\ }\textbf {\bibinfo {volume} {48}},\ \bibinfo
  {pages} {4460} (\bibinfo {year} {2023})}\BibitemShut {NoStop}%
\bibitem [{\citenamefont {Zhao}\ \emph {et~al.}(2019)\citenamefont {Zhao},
  \citenamefont {Guo},\ and\ \citenamefont {Fan}}]{Zhao19}%
  \BibitemOpen
  \bibfield  {author} {\bibinfo {author} {\bibfnamefont {Z.}~\bibnamefont
  {Zhao}}, \bibinfo {author} {\bibfnamefont {C.}~\bibnamefont {Guo}},\ and\
  \bibinfo {author} {\bibfnamefont {S.}~\bibnamefont {Fan}},\ }\bibfield
  {title} {\bibinfo {title} {Connection of temporal coupled-mode-theory
  formalisms for a resonant optical system and its time-reversal conjugate},\
  }\href {https://doi.org/10.1103/physreva.99.033839} {\bibfield  {journal}
  {\bibinfo  {journal} {Physical Review A}\ }\textbf {\bibinfo {volume} {99}},\
  \bibinfo {pages} {033839} (\bibinfo {year} {2019})}\BibitemShut {NoStop}%
\bibitem [{\citenamefont {Breiman}(2001)}]{Breiman2001}%
  \BibitemOpen
  \bibfield  {author} {\bibinfo {author} {\bibfnamefont {L.}~\bibnamefont
  {Breiman}},\ }\bibfield  {title} {\bibinfo {title} {Random forests},\ }\href
  {https://doi.org/10.1023/A:1010933404324} {\bibfield  {journal} {\bibinfo
  {journal} {Machine Learning}\ }\textbf {\bibinfo {volume} {45}},\ \bibinfo
  {pages} {5} (\bibinfo {year} {2001})}\BibitemShut {NoStop}%
\bibitem [{\citenamefont {Ho}(1995)}]{Ho1995}%
  \BibitemOpen
  \bibfield  {author} {\bibinfo {author} {\bibfnamefont {T.~K.}\ \bibnamefont
  {Ho}},\ }\bibfield  {title} {\bibinfo {title} {Random decision forests},\
  }in\ \href {https://doi.org/10.1109/ICDAR.1995.598994} {\emph {\bibinfo
  {booktitle} {Proceedings of 3rd International Conference on Document Analysis
  and Recognition}}},\ Vol.~\bibinfo {volume} {1}\ (\bibinfo {year} {1995})\
  pp.\ \bibinfo {pages} {278--282 vol.1}\BibitemShut {NoStop}%
\bibitem [{\citenamefont {Liu}\ \emph {et~al.}(2012)\citenamefont {Liu},
  \citenamefont {Wang},\ and\ \citenamefont {Zhang}}]{Liu2012}%
  \BibitemOpen
  \bibfield  {author} {\bibinfo {author} {\bibfnamefont {Y.}~\bibnamefont
  {Liu}}, \bibinfo {author} {\bibfnamefont {Y.}~\bibnamefont {Wang}},\ and\
  \bibinfo {author} {\bibfnamefont {J.}~\bibnamefont {Zhang}},\ }\bibinfo
  {title} {New machine learning algorithm: Random forest},\ in\ \href
  {https://doi.org/10.1007/978-3-642-34062-8_32} {\emph {\bibinfo {booktitle}
  {Information Computing and Applications}}}\ (\bibinfo  {publisher} {Springer
  Berlin Heidelberg},\ \bibinfo {year} {2012})\ p.\ \bibinfo {pages}
  {246–252}\BibitemShut {NoStop}%
\bibitem [{\citenamefont {Segal}()}]{Segal1993}%
  \BibitemOpen
  \bibfield  {author} {\bibinfo {author} {\bibfnamefont {M.~R.}\ \bibnamefont
  {Segal}},\ }\href@noop {} {\bibinfo {title} {Machine learning benchmarks and
  random forest regression}},\ \bibinfo {howpublished}
  {\url{https://escholarship.org/uc/item/35x3v9t4}}\BibitemShut {NoStop}%
\bibitem [{\citenamefont {Van~Rossum}\ and\ \citenamefont {{Python Dev
  Team}}(2016)}]{Van_Rossum2016-yc}%
  \BibitemOpen
  \bibfield  {author} {\bibinfo {author} {\bibfnamefont {G.}~\bibnamefont
  {Van~Rossum}}\ and\ \bibinfo {author} {\bibnamefont {{Python Dev Team}}},\
  }\href@noop {} {\emph {\bibinfo {title} {Python 3.6 Language Reference}}}\
  (\bibinfo  {publisher} {Samurai Media},\ \bibinfo {year} {2016})\BibitemShut
  {NoStop}%
\bibitem [{\citenamefont {Altmann}\ \emph {et~al.}(2010)\citenamefont
  {Altmann}, \citenamefont {Toloşi}, \citenamefont {Sander},\ and\
  \citenamefont {Lengauer}}]{Altmann2010}%
  \BibitemOpen
  \bibfield  {author} {\bibinfo {author} {\bibfnamefont {A.}~\bibnamefont
  {Altmann}}, \bibinfo {author} {\bibfnamefont {L.}~\bibnamefont {Toloşi}},
  \bibinfo {author} {\bibfnamefont {O.}~\bibnamefont {Sander}},\ and\ \bibinfo
  {author} {\bibfnamefont {T.}~\bibnamefont {Lengauer}},\ }\bibfield  {title}
  {\bibinfo {title} {Permutation importance: a corrected feature importance
  measure},\ }\href {https://doi.org/10.1093/bioinformatics/btq134} {\bibfield
  {journal} {\bibinfo  {journal} {Bioinformatics}\ }\textbf {\bibinfo {volume}
  {26}},\ \bibinfo {pages} {1340–1347} (\bibinfo {year} {2010})}\BibitemShut
  {NoStop}%
\bibitem [{\citenamefont {Wehenkel}\ \emph {et~al.}(2018)\citenamefont
  {Wehenkel}, \citenamefont {Sutera}, \citenamefont {Bastin}, \citenamefont
  {Geurts},\ and\ \citenamefont {Phillips}}]{Wehenkel2018}%
  \BibitemOpen
  \bibfield  {author} {\bibinfo {author} {\bibfnamefont {M.}~\bibnamefont
  {Wehenkel}}, \bibinfo {author} {\bibfnamefont {A.}~\bibnamefont {Sutera}},
  \bibinfo {author} {\bibfnamefont {C.}~\bibnamefont {Bastin}}, \bibinfo
  {author} {\bibfnamefont {P.}~\bibnamefont {Geurts}},\ and\ \bibinfo {author}
  {\bibfnamefont {C.}~\bibnamefont {Phillips}},\ }\bibfield  {title} {\bibinfo
  {title} {Random forests based group importance scores and their statistical
  interpretation: Application for alzheimer’s disease},\ }\bibfield
  {journal} {\bibinfo  {journal} {Frontiers in Neuroscience}\ }\textbf
  {\bibinfo {volume} {12}},\ \href {https://doi.org/10.3389/fnins.2018.00411}
  {10.3389/fnins.2018.00411} (\bibinfo {year} {2018})\BibitemShut {NoStop}%
\bibitem [{\citenamefont {Gippius}\ \emph {et~al.}(2005)\citenamefont
  {Gippius}, \citenamefont {Tikhodeev},\ and\ \citenamefont
  {Ishihara}}]{Gippius2005}%
  \BibitemOpen
  \bibfield  {author} {\bibinfo {author} {\bibfnamefont {N.~A.}\ \bibnamefont
  {Gippius}}, \bibinfo {author} {\bibfnamefont {S.~G.}\ \bibnamefont
  {Tikhodeev}},\ and\ \bibinfo {author} {\bibfnamefont {T.}~\bibnamefont
  {Ishihara}},\ }\bibfield  {title} {\bibinfo {title} {Optical properties of
  photonic crystal slabs with an asymmetrical unit cell},\ }\href
  {https://doi.org/10.1103/physrevb.72.045138} {\bibfield  {journal} {\bibinfo
  {journal} {Physical Review B}\ }\textbf {\bibinfo {volume} {72}},\ \bibinfo
  {pages} {045138} (\bibinfo {year} {2005})}\BibitemShut {NoStop}%
\end{thebibliography}
\end{document}